\documentclass[usenatbib]{mn2e}

\usepackage[dvips]{graphicx}
\usepackage{amssymb}
\usepackage{txfonts}
\usepackage{colordvi}

\newcommand{\integral}{{\textit{INTEGRAL}}}
\newcommand{\xte}{{\textit{RXTE}}}

\newcommand{\sax}{{\textit{Beppo\-SAX}}}
\newcommand{\gro}{{\textit{CGRO}}}
\newcommand{\fermi}{{\textit{Fermi}}}
\newcommand{\agile}{{\textit{AGILE}}}

\newcommand{\msun}{{\rm M}_{\sun}}

\newcommand{\g}{$\gamma$}

\newbox\grsign \setbox\grsign=\hbox{$>$} \newdimen\grdimen \grdimen=\ht\grsign
\newbox\simlessbox \newbox\simgreatbox \newbox\simpropbox
\setbox\simgreatbox=\hbox{\raise.5ex\hbox{$>$}\llap
     {\lower.5ex\hbox{$\sim$}}}\ht1=\grdimen\dp1=0pt
\setbox\simlessbox=\hbox{\raise.5ex\hbox{$<$}\llap
     {\lower.5ex\hbox{$\sim$}}}\ht2=\grdimen\dp2=0pt
\setbox\simpropbox=\hbox{\raise.5ex\hbox{$\propto$}\llap
     {\lower.5ex\hbox{$\sim$}}}\ht2=\grdimen\dp2=0pt
\def\ga{\mathrel{\copy\simgreatbox}}
\def\la{\mathrel{\copy\simlessbox}}

\title[The MeV tail in Cyg X-1]{The MeV spectral tail in Cyg X-1 and optically-thin emission of jets}

\author[A. A. Zdziarski et al.]
{Andrzej A. Zdziarski,$^1$\thanks{E-mail: aaz@camk.edu.pl} Piotr Lubi\'nski$^2$  and Marek Sikora$^1$\\
$^1$Centrum Astronomiczne im.\ M. Kopernika, Bartycka 18, PL-00-716 Warszawa, Poland\\
$^2$Centrum Astronomiczne im.\ M. Kopernika, Rabia\'nska 8, PL-87-100 Toru\'n, Poland\\
}

\date{Accepted 2012 March 12. Received 2012 March 12; in original form 2012 January 07}

\pagerange{\pageref{firstpage}--\pageref{lastpage}}
\pubyear{2012}

\begin{document}

\maketitle

\label{firstpage}

\begin{abstract}
We study the average X-ray and soft \g-ray spectrum of Cyg X-1 in the hard spectral state, using data from \integral. We compare these results with those from \gro, and find a good agreement. Confirming previous studies, we find the presence of a high-energy MeV tail beyond a thermal-Comptonization spectrum; however, the tail is much softer and weaker than that recently published by Laurent et al. In spite of this difference, the observed high-energy tail could still be due to the synchrotron emission of the jet of Cyg X-1, as claimed by Laurent et al.

In order to test this possibility, we study optically-thin synchrotron and self-Compton emission from partially self-absorbed jets. We develop formalisms for calculating both emission of the jet base (which we define here as the region where the jet starts its emission) and emission of the entire jet. We require the emission to match that observed at the turnover energy. The optically thin emission is dominated by that from the jet base, and it has to become self-absorbed within it at the turnover frequency. We find this implies the magnetic field strength at the jet base of $B_0\propto z_0^4$, where $z_0$ is the distance of the base from the black-hole centre. The value of $B_0$ is then constrained from below by the condition that the self-Compton emission is below an upper limit in the GeV range, and from above by the condition that the Poynting flux does not exceed the jet kinetic power. This yields $B_0$ of the order of $\sim 10^4$ G and the location of the jet base at $\sim 10^3$ gravitational radii. Using our formalism, we find the MeV tail can be due to jet synchrotron emission, but this requires the electron acceleration at a rather hard power-law index, $p\simeq 1.3$--1.6. For acceleration indices of $p\ga 2$, the amplitude of the synchrotron component is much below that of MeV tail, and its origin is likely to be due to hybrid Comptonization in the accretion flow.
\end{abstract}

\section{Introduction}
\label{intro}

\begin{figure*}
\centerline{\includegraphics[width=16cm]{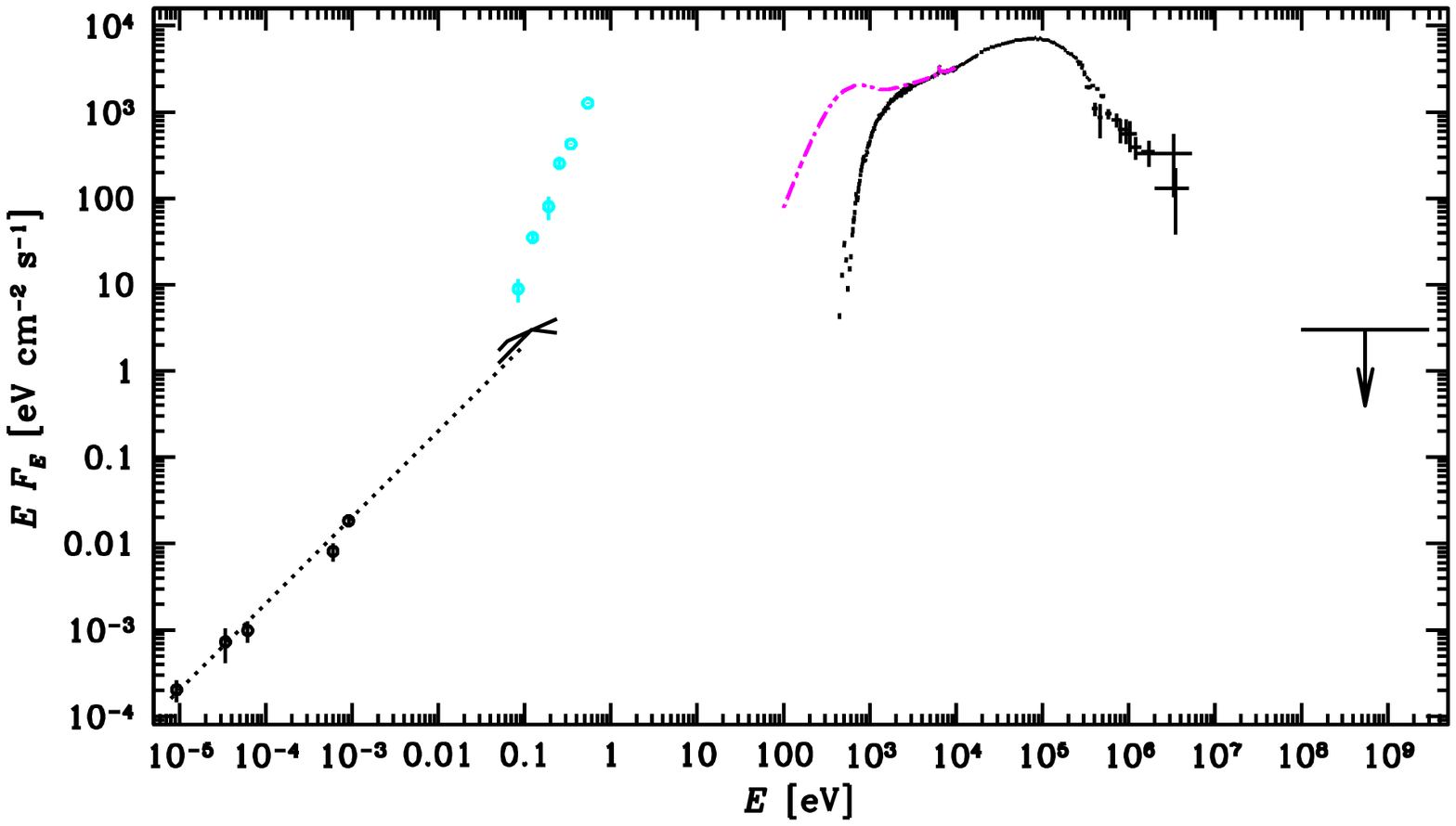}} 
\caption{The hard-state radio to \g-ray spectrum of Cyg X-1. The five radio/mm fluxes are from \citet{fender00}. The two IR broken power-law spectra show the jet component for observations 1 and 2 of R11. The dotted line shows an $\alpha=0$ radio spectrum extrapolated up to 0.1 eV. The cyan circles and error bars show the total IR fluxes \citep{persi80,mirabel96}, which are mostly from emission of the companion and its wind. The black X-ray and soft \g-ray data points above 20 keV show the IBIS and COMPTEL spectra. The spectrum below 20 keV, from \sax, represents a typical hard state spectrum (which is absorbed by an intervening medium). The dot-dashed magenta curve shows a fitted model of the unabsorbed spectrum. The \g-ray upper limit is from \agile\/ \citep{sabatini10}. See Sections \ref{pars}--\ref{spectra} for details. 
} \label{r_x_gamma}
\end{figure*}

Cyg X-1 is an archetypical and well-studied black-hole binary. In the hard spectral state, its hard X-ray/soft \g-ray spectrum has been known to be hard below the maximum [in the $EF(E)$ representation] at $\sim$100--200 keV. Above it, the spectrum shows a sharp cutoff, which then is followed by a high-energy tail, measured up to a few MeV (\citealt{mcconnell02}, hereafter M02). 

We consider here the finding of \citet{l11} (hereafter L11) of the tail at 0.4--2 MeV being very strong and hard, with the energy spectral index of $\alpha\simeq 0.6\pm 0.2$. The tail measured by L11 was found to be strongly polarized, which would imply it is due to synchrotron emission of the jet in this system. We calculate the average hard-state \integral\/ spectra from the ISGRI and PICsIT detectors, and although we do find a high-energy tail above about the same energy as L11, it is weaker by about an order of magnitude as well as much softer, with $\alpha\simeq 2$. The weakness of the tail is also confirmed by an independent study of the average \integral\/ SPI spectrum \citep*{jrm12}, and it is in agreement with the \gro\/ results of M02.

As known before, the high-energy tail can be well fitted by emission of non-thermal electrons forming a high-energy tail beyond the Maxwellian distribution, responsible for the bulk of the X-ray emission. On the other hand, the overall spectrum can also be fitted by purely thermal Comptonization (presumably in a hot accretion flow) and emission of power-law electrons with an exponential cutoff, which electrons could be located within the jet.

In order to test the jet origin of the tail, we study emission of partially self-absorbed jets, such as that present in the hard state of Cyg X-1. We develop two formalisms. In one, a one-zone model of the jet base is developed. In the other, we model the emission of the entire jet (reformulating the model of \citealt{bk79}, hereafter BK79). We find that the optically-thin emission is dominated by the jet base. In both approaches, we relate the magnetic field at the jet base to its height along the jet. We apply our model to Cyg X-1, using available measurements and upper limits. We find that the observed turnover energy, the flux at that energy, the MeV tail measurements, an upper limit on the emission in the GeV region, and a measurement of the kinetic jet power allow a relatively precise determination of both the height of the jet base and its magnetic field.

\section{Notation and the parameters of Cyg X-1}
\label{pars}

Hereafter, we use the following dimensionless parameters,
\begin{equation}
\epsilon\equiv {E'\over m_{\rm e}c^2},\quad \zeta\equiv {z\over R_{\rm g}},\quad \dot m\equiv{\dot M c^2\over L_{\rm E}},\quad p_{\rm j}\equiv{P_{\rm j} \over L_{\rm E}},
\label{units}
\end{equation}
where $E'$ is the photon energy in the jet comoving frame, $z$ is the height along a jet (from the centre of the black hole in the observer's frame), $R_{\rm g}= GM/c^2$ is the gravitational radius, $G$ is the gravitational constant, $M$ is the black hole mass, $\dot M$ is the mass accretion rate, $P_{\rm j}$ is the sum of the total kinetic powers of the jet and counterjet, $L_{\rm E}=4\upi G M m_{\rm p}c/\sigma_{\rm T}$ is the Eddington luminosity, $\sigma_{\rm T}$ is the Thomson cross section, and $m_{\rm e}$, $m_{\rm p}$ and are the electron and proton mass, respectively. We use then the symbol $E$ for the observed photon energy.

We adopt the best-fit values of \citet{orosz11} of $M\simeq 15\msun$ (where $\msun$ is the solar mass) and the inclination of the normal to the binary plane with respect to the line of sight of $i=27\degr$. We assume that the jet and the disc have the same inclination. We use the best-fit value of the distance to Cyg X-1 of \citet{reid11} of $D=1.86$ kpc. The separation between the stars is $a\simeq 3\times 10^{12}$ cm, and the stellar luminosity is $L_*\simeq 10^{39}$ erg s$^{-1}$ \citep{cn09,orosz11}. 

The opening angle of the jet in Cyg X-1 on the length scale of $\sim 10^{15}$ cm has been constrained by VLBA and VLA observations to $\Theta_{\rm j}\la 2\degr$ \citep{stirling01}, and its velocity, to $\beta_{\rm j}\simeq 0.6$ \citep*{stirling01,gleissner04,mbf09}, which corresponds to the Lorentz factor of $\Gamma_{\rm j}=(1-\beta_{\rm j}^2)^{-1/2}\simeq 1.25$. The kinetic jet power in the hard state of Cyg X-1 has been estimated as $P_{\rm j}\simeq (1$--$3)\times 10^{37}$ erg s$^{-1}$ from the optical nebula presumably powered by the jet \citep{gallo05,russell07}. For $M=15\msun$, $L_{\rm E}\simeq 2\times 10^{39}$ erg s$^{-1}$, which gives $p_{\rm j}\sim 10^{-2}$. On the other hand, $\dot m\simeq 0.1$ is obtained for the observed average X-ray luminosity of $L_{\rm X}\simeq 10^{37}$ erg s$^{-1}$ in the hard state of Cyg X-1 \citep{z02} using an accretion efficiency of 0.05. 

The position angle of the radio jet on mas scale found by \citet{stirling01} (see also \citealt{rushton11}) is $-(17\degr$--$24\degr)$. However, the sign of these values was misprinted as positive in \citet{stirling01}. The actual position angle, conventionally measured from the north to the east, is negative, as stated, e.g., in \citet{spencer01}.

A broad-band spectrum of Cyg X-1 in the hard state is shown in Fig.\ \ref{r_x_gamma}. In the radio/mm range, we show the fluxes at 2.3, 8.3, 15, 146 and 220 GHz (from Table 1 of \citealt{fender00}). These fluxes can be joined by an $\alpha=0$ spectrum, which extends up to the turnover energy at $E_{\rm t}\simeq 0.1$ eV in the IR. The two 0.05--0.24 eV broken power-law spectra show the fitted jet component for observations 1 and 2 (case 2, model c) of \citet{rahoui11} (hereafter R11), with the fitted values of $E_{\rm t}\simeq 0.12$ eV and 0.06 eV, respectively. The cyan crosses and error bars show the measurements of \citet{persi80} and \citet{mirabel96} of the total IR emission, which is dominated by the companion and its wind. Note that the fitted jet component represents a relatively small fraction of the total IR emission, especially at its high-energy end. The black 20 keV--5 MeV data points show the average ISGRI and PICsIT spectra obtained by us (see below), and the average 0.7--5 MeV COMPTEL spectrum from M02. The spectrum shown at 0.5--20 keV is an example of a typical hard-state spectrum, observed by \sax\/ \citep{ds01}. Note that this spectrum is strongly absorbed at low energies by an intervening medium; the magenta dot-dashed curve shows a model fit of the intrinsic spectrum. The 0.1--3 GeV upper limit is from \agile\/ \citep{sabatini10}, with the integral photon flux of $<3\times 10^{-8}$ cm$^{-2}$ s$^{-1}$ ($2\sigma$). Given the large width of the energy band with a strong dependence of an implied monochromatic upper limit on the spectral shape, we use here the same $EF(E)$ limit as that of fig.\ 3 of \citet{sabatini10}. This approximately corresponds to the monochromatic flux of $\la F_\gamma= 6\times 10^{-9}$ cm$^{-2}$ s$^{-1}$ in the middle of the observed range at $E_\gamma=0.5$ GeV. The present upper limit is a factor of a few below that of the \gro/EGRET (given in \citealt{zg04}). 

\section{The X-ray and soft \mbox{\boldmath${\gamma}$}-ray spectra}
\label{spectra}

\subsection{Analysis of the \textit{INTEGRAL}\/ data}
\label{data}

We study \integral\/ data from the IBIS instrument. We consider separately its two detectors, the ISGRI and PICsIT. We use data from the \integral\/ revolutions from 22 to 877 (MJD 52626--55186, the years 2003--2009), which are available in the \integral\/ Science Data Centre (ISDC). All public data with the off-axis angle $< 15\degr$ were selected (a $< 9\degr$ condition gave virtually the same spectra). The ISGRI and PICsIT data are used in the 19--500 keV and 0.25--5.4 MeV ranges, respectively. 

We obtain both the average spectrum from all the above data, and the average spectrum in the hard spectral state. The former is still dominated by the hard spectral state, which is due to Cyg X-1 having been mostly in it since the launch of \integral\/ until 2010 June. Thus, the differences between the two sets are relatively minor. As the hard-state intervals, we take MJD 50350--50590, 50660--50995, 51025--51400, 51640--51840, 51960--52100, 52565--52770, 52880--52975, 53115--53174, 53540--53800, 53880--55375, following \citet{zdz12}. In addition, we also use the data from MJD 53188--53215, during which the spectra appeared typical to the hard state. 

Our data analysis procedure is as follows. The ISGRI data have been reduced using the Offline Scientific Analysis ({\sc osa}) 9.0 provided by the ISDC \citep{Courvoisier2003}, with the pipeline parameters set to the default values. For the spectral extraction we took into account also 22 strongest ISGRI sources in the Cyg X-1 neighbourhood. The PICsIT spectra were prepared using a non-standard software, optimized for handling Poisson-distributed data affected by a strong and variable background \citep{Lubinski2009}. We have added a 1 per cent systematic error to each of the ISGRI and PICsIT data sets.

\subsection{Average spectra}
\label{spectral}

\begin{figure}
\centerline{\includegraphics[width=\columnwidth]{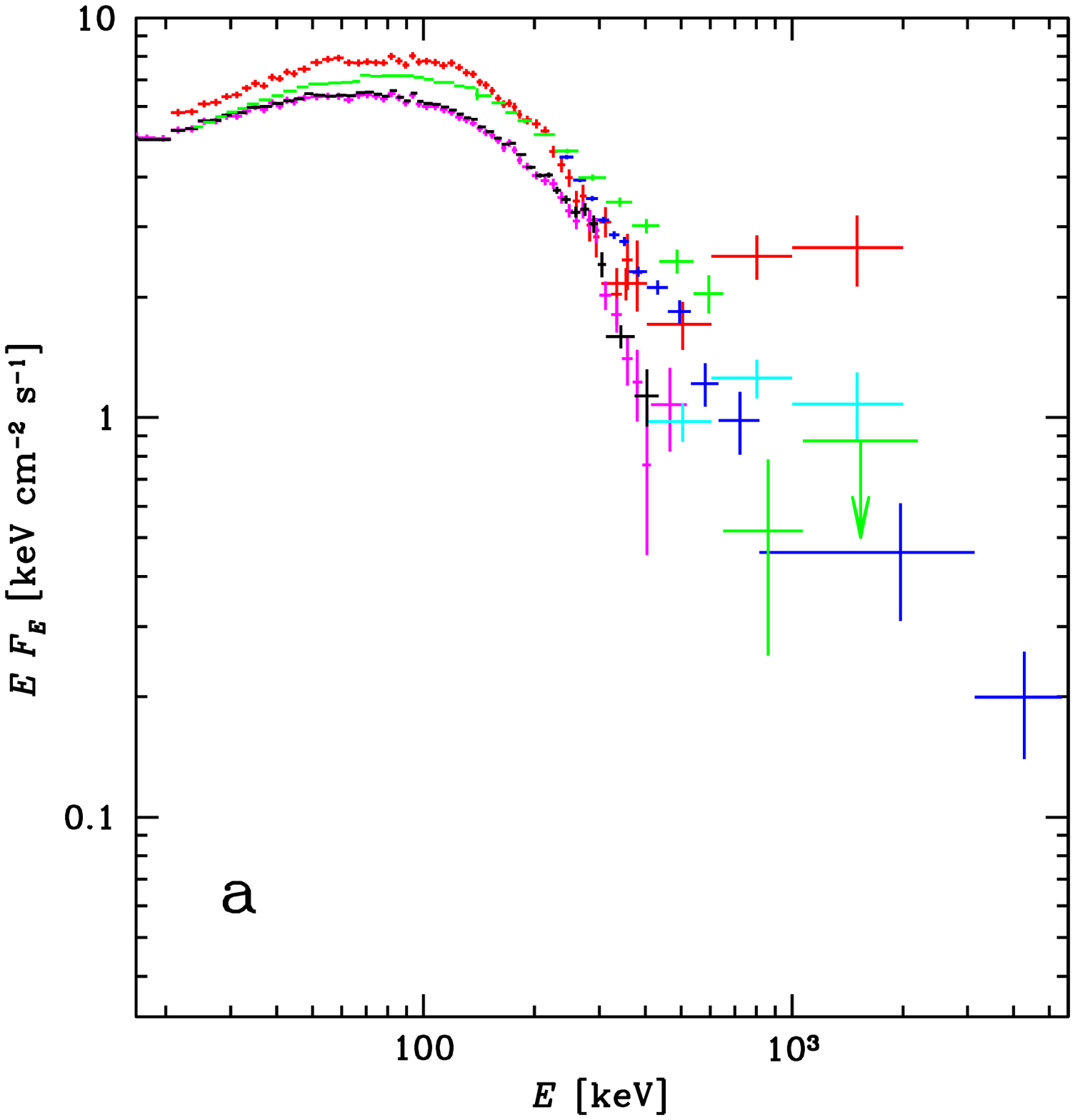}}
\centerline{\includegraphics[width=\columnwidth]{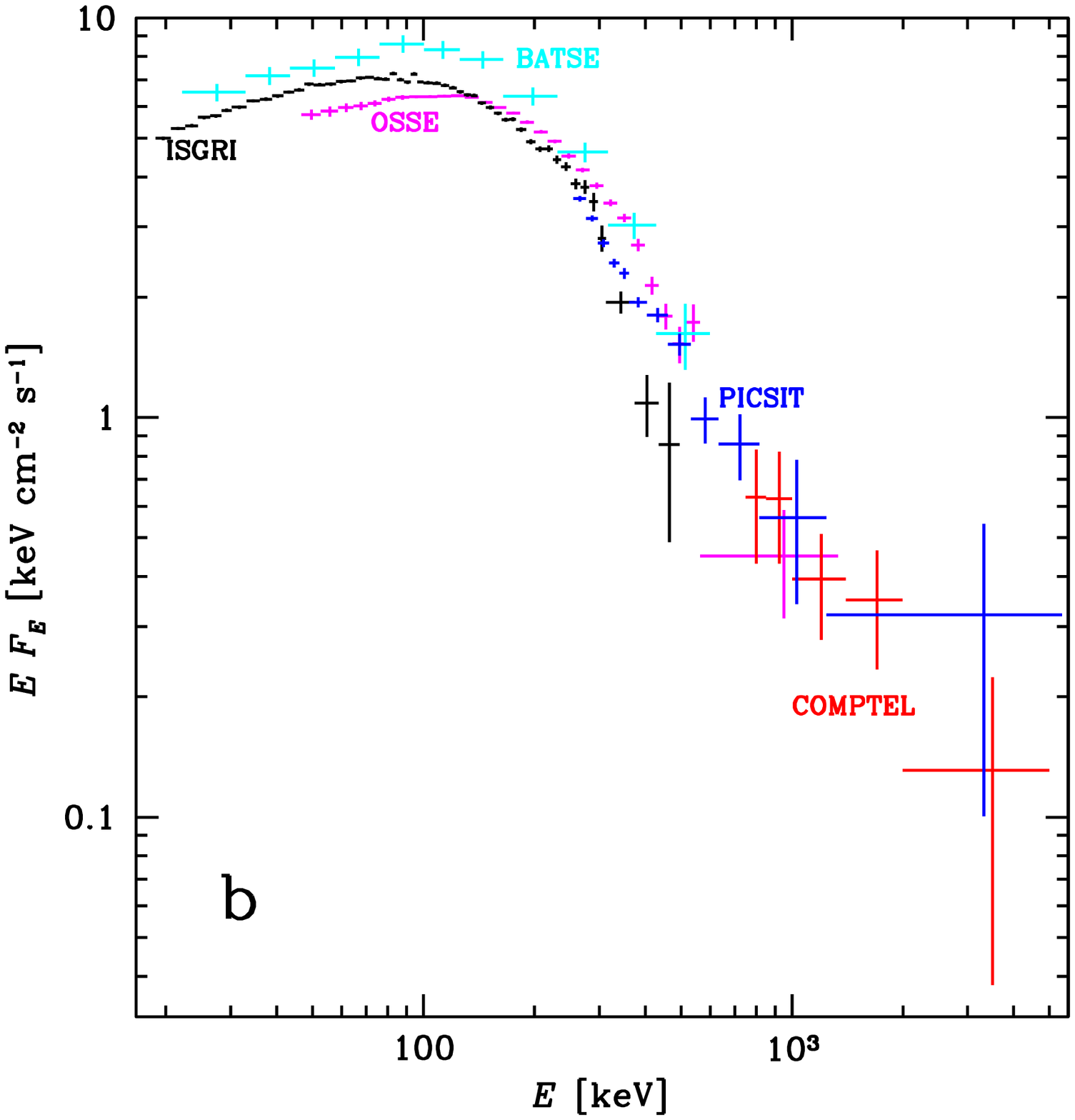}}
\caption{Comparison of average 20 keV--5 MeV spectra of Cyg X-1. (a) The spectra averaged over all \integral\/ observations from 2003--2009. The black and blue spectra are from our analysis of the data from the ISGRI and PICsIT, respectively, fitted together by hybrid Comptonization. The red spectra are from the ISGRI and the Compton mode (ISGRI+PICsIT), as published by L11. The magenta and cyan crosses are the corrected ISGRI and Compton-mode spectra, respectively (Ph.\ Laurent, private communication). Each of those two pairs of spectra have been fitted by thermal Comptonization and a power law. The green spectra are from the SPI, fitted by two thermal-Compton components \citep{jrm12}. (b) The average spectra in the hard spectral state from \integral\/ and \gro. The black and blue spectra are from the ISGRI and PICsIT (with the PICsIT data renormalized to the level of the ISGRI). The magenta, cyan and red symbols show the \gro\/ OSSE, BATSE and COMPTEL spectra, respectively, fitted together by hybrid Comptonization (M02). See Section \ref{spectra} for details. 
}
\label{f:spectra}
\end{figure}

\begin{figure}
\centerline{\includegraphics[width=\columnwidth]{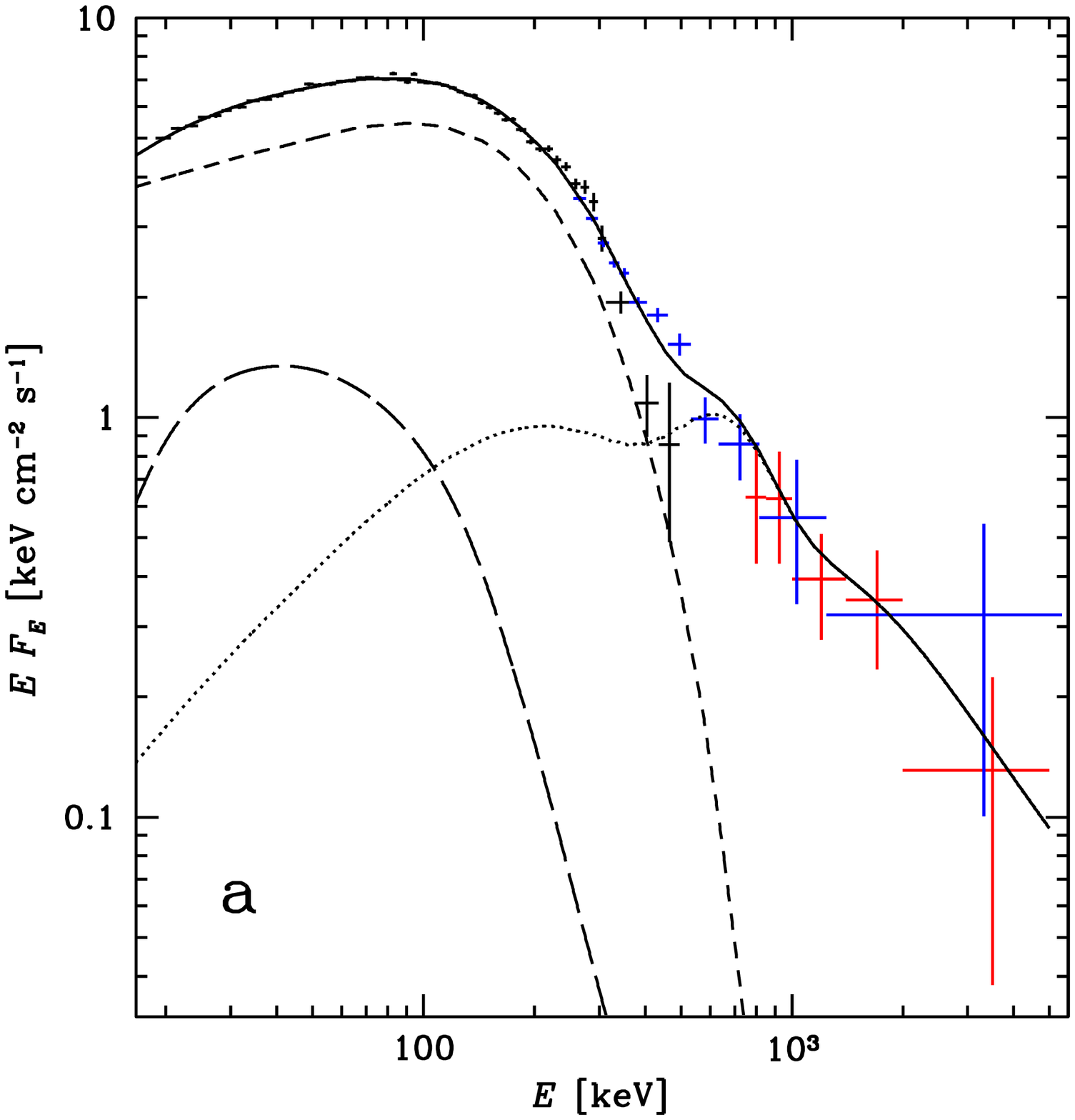}}
\centerline{\includegraphics[width=\columnwidth]{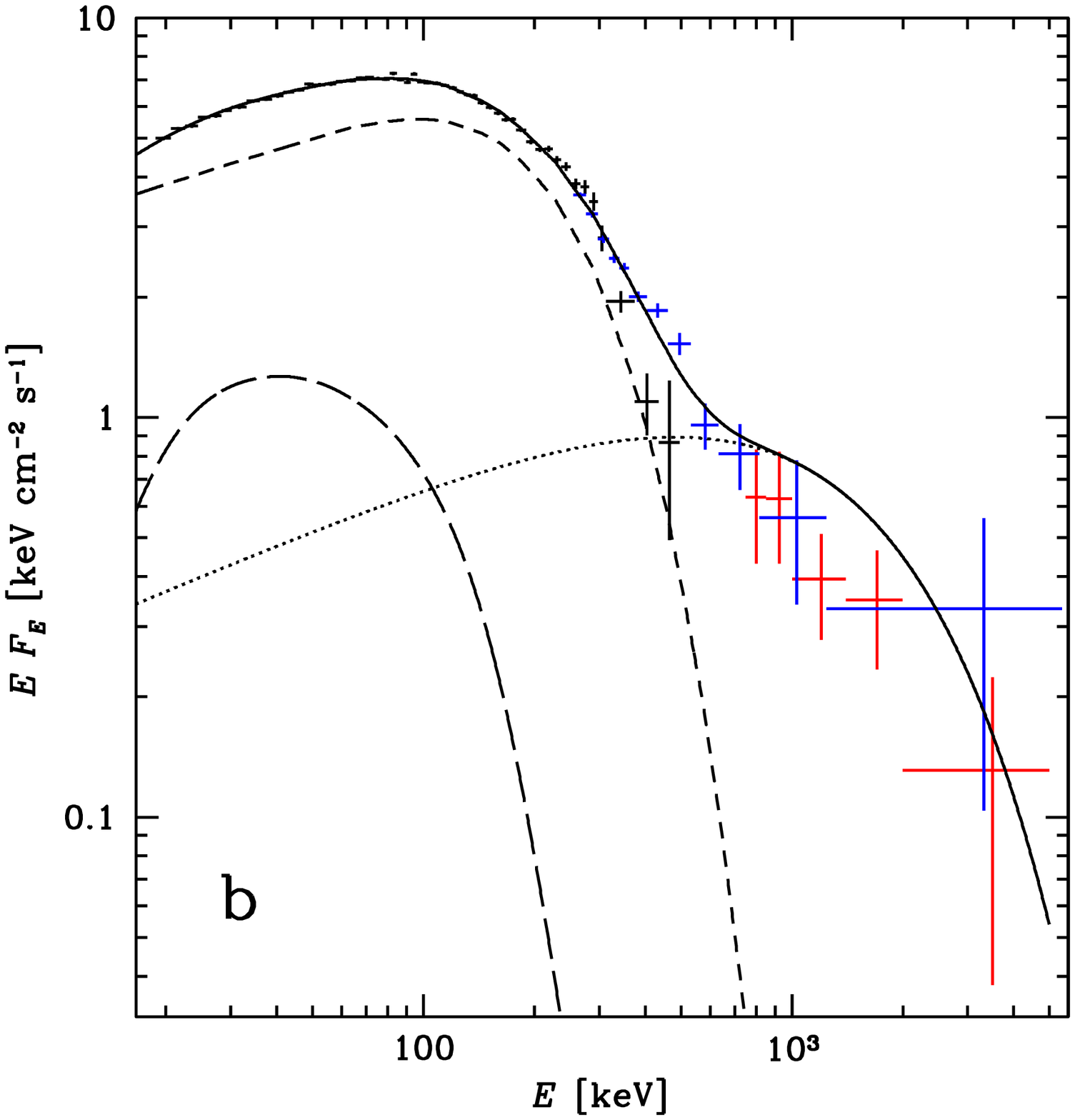}}
\caption{The average 20 keV--5 MeV spectra of Cyg X-1 from our analysis of the ISGRI (black) and PICsIT (blue) in the hard spectral state fitted by (a) hybrid Comptonization and reflection and (b) thermal Comptonization, reflection, and an e-folded power law. For comparison, the COMPTEL spectrum from M02 (not taken into account in the fits) is shown in red. The short and long dashes and the solid curves show the thermal Compton scattering, Compton reflection, and the total model, respectively. The dotted curve in (a) shows the Compton emission from the non-thermal e$^\pm$ plus the annihilation spectrum from the e$^\pm$ pairs, and in (b), the e-folded power-law component. See Section \ref{spectra} for details. 
}
\label{f:2c}
\end{figure}

We have fitted our IBIS spectra for both all the observations and those in the hard state only with the model of hybrid Comptonization, {\sc eqpair}, obtaining results similar to those of M02. The deconvolved spectra for the two data sets are shown in Figs.\ \ref{f:spectra}(a) and (b), respectively. Since the fit results are similar for both data sets, and since the former contains also some contribution from soft/intermediate states, we describe the fit results only for the latter below. We note that the overall flux calibration for the PICsIT differs slightly from that for the ISGRI, yielding the PICsIT flux about 1.3 times higher than that of ISGRI in the overlapping energy range. In Fig.\ \ref{f:spectra}(a), the PICsIT spectrum is shown at its original normalization, whereas in Figs.\ \ref{f:spectra}(b)--\ref{f:2c} it is plotted renormalized to the level of ISGRI. We note that such differences between the normalization of different instruments are common, e.g., a similar difference was present for the PCA and HEXTE detectors onboard \xte.

Fig.\ \ref{f:spectra}(a) compares then our IBIS spectrum with that published in L11. The latter spectrum is from the ISGRI up to $\sim$400 keV, and at higher energies from the Compton mode, which utilizes both ISGRI and PICsIT. We see that those fluxes are substantially above ours, by about 30 per cent at 100 keV, and up to a factor of several above 1 MeV. However, it turns out (Ph.\ Laurent, private communication) that the published spectrum does not represent the average Cyg X-1 spectrum from the IBIS, and it does not correspond to the data used to calculate the polarization. The corrected average spectrum (Ph.\ Laurent, private communication) is shown in magenta and cyan in Fig.\ \ref{f:spectra}(a). Its ISGRI part agrees with our ISGRI spectrum almost perfectly. The Compton-mode spectrum still shows a disagreement with our PICsIT spectrum at $\ga 1$ MeV, but now it is at a level of $\sim$(2--$3)\sigma$, compared to several $\sigma$ discrepancy for the published spectrum.

In Fig.\ \ref{f:spectra}(a), we also show the average spectrum from the \integral\/ SPI detector of \citet{jrm12}. This spectrum was obtained using all usable SPI Cyg X-1 data, which time interval spans 52797--55184, and which, similarly to our average, also includes some fraction of soft/intermediate-state data. Although there is an overall agreement with our results, the SPI spectrum is above ours at $E\ga 30$ keV, and shows a much harder decline up to 700 keV. This may be partly due to the ISGRI calibration with respect to the Crab spectrum, assuming that spectrum above 100 keV has $\alpha=1.35$, whereas current estimates give $\alpha=1.24\pm 0.02$ \citep{jr09}. This also seems to explain our PICsIT spectrum being harder than that from ISGRI at $\ga 150$ keV. We also see that the $2\sigma$ upper limit from SPI at 1.1--2.2 MeV is above our PICsIT flux but still somewhat below the corrected Compton-mode flux (of Ph.\ Laurent, private communication).

Fig.\ \ref{f:spectra}(b) compares then our \integral/IBIS average spectra for the hard state with the averaged ones for 1991--1994 measured by \gro\/ (M02). Given uncertainties in the instrument calibration, we have not applied any correcting factors to the BATSE and COMPTEL data, which were applied in M02 in order to normalize those data to the OSSE spectrum. We see that our IBIS spectrum is very similar to that from BATSE except for a slight difference in the normalization. At $E\ga 0.4$ MeV, the PICsIT, COMPTEL, BATSE and OSSE spectra all agree well. 

The {\sc eqpair} model is described in detail in \citet{c99} and \citet{gierlinski99}. Our best-fit parameters for the hard state are: the ratio of the power supplied to the Comptonizing plasma to that in the seed photons irradiating it of $6.8\pm 0.2$, the fraction of the power supplied to the non-thermal electrons in the plasma of $0.16\pm 0.01$, the Thomson optical depth of ionization electrons of $\tau_{\rm i}=0.79\pm 0.07$, the relative strength of reflection (using the method of \citealt{mz95}) of $\Omega/2\upi=0.25\pm 0.03$. There is a relatively strong e$^\pm$ pair production in this model, with the total Thomson optical depth including pairs of 1.44, and at the equilibrium temperature of e$^\pm$ of $kT_{\rm e}\simeq 69$ keV (note that these two quantities are not parameters of the model but instead result from the fitted and assumed parameters). We have assumed the compactness parameter of 10 of the blackbody seed photons at the temperature of 0.2 keV. The amount of pair production depends on the former, and the dependence on the latter is rather weak, see discussion in \citet{gierlinski99}. The model and its components, i.e., scattering by thermal electrons, scattering by non-thermal electrons plus pair annihilation (which is due to pairs produced by the non-thermal part of the spectrum), and Compton reflection, are shown in Fig.\ \ref{f:2c}(a). An annihilation hump is seen in the second component. The decomposition of the Compton spectrum into the thermal and non-thermal components has been done using the method described in \citet{hannikainen05}. We have assumed that the model has a 1 per cent systematic error. We have obtained a relatively good fit, $\chi^2/{\rm dof}=94/58$. The model is shown by the solid curve in Fig.\ \ref{f:spectra}(b). Note that although we have not fitted the COMPTEL data, our model agrees well with them. 

We have then tested a two-component model in which the main part of the spectrum was fitted by thermal Comptonization without any non-thermal tail, and the observed tail was attributed to a power law with an exponential cutoff. We have found that the energy index, $\alpha$, of the power law could not be constrained as the same $\chi^2$ was found in a wide range of $\alpha$. However, since a possible physical interpretation of this component is a synchrotron contribution from the jet, we have fixed it at $\alpha=0.6$, for which this power law extrapolated down to the turnover energy of $E_{\rm t}\simeq 0.1$ eV connects to the extrapolation of the optically thick jet emission with $\alpha\simeq 0$, see Fig.\ \ref{r_x_gamma}. We have obtained the e-folding energy of $E_{\rm m}=1.2\pm 0.3$ MeV and the 1-keV normalization of 0.11 cm$^{-2}$ s$^{-1}$, at $\chi^2/{\rm dof}=96/57$, i.e., only slightly worse than that of the hybrid-Compton model. The thermal plasma parameters are: the power ratio of $6.5\pm 0.2$, $\tau_{\rm i}=1.52\pm 0.03$ (with virtually no pair production, and with $kT_{\rm e}=69$ keV), and $\Omega/2\upi= 0.26\pm 0.03$. We see that the two-component model predicts a hump at $\sim$2 MeV somewhat more pronounced than the hybrid-Compton model. The two components intersect at 410 keV. Thus, if the e-folded power law were strongly polarized, it could explain the polarization measured above 400 keV by L11.

\section{Models of jet emission}
\label{jet}

We study both a one-zone model of the jet base region and a jet model of the entire jet, and apply it to Cyg X-1. For our purposes, we define the jet base as the location at which the jet emission starts (assuming continuous jet emission, as in the model of BK79). This is likely to differ from the actual base where the jet is formed but the dissipation has not yet started. 

As we confirm in Section \ref{self-abs}, the base provides most of the optically-thin jet emission. The jet base becomes optically thin at the turnover energy, $E_{\rm t}$. The value of this energy and the synchrotron flux at $E_{\rm t}$ in Cyg X-1 have been measured by R11. This allows us to determine the normalization of the distribution of the emitting electrons as a function of the magnetic field. The synchrotron emission has to be self-absorbed within the jet base at $E_{\rm t}$, which imposes a relationship between the magnetic field, $B_0$, and the location of the base, $z_0$. Then the magnetic field strength is constrained from below by the requirement that the self-Compton emission in the \g-ray energy range does not exceed the observational upper limit. On the other hand, the magnetic energy flux cannot be greater than the observed jet kinetic energy. 

\subsection{One-zone model of the jet base}
\label{base}

The radio up to far IR emission of Cyg X-1 is flat in the ${\rm d}F/ {\rm d}E$ representation, $\propto E^\alpha$, with $\alpha\simeq 0$ (see Fig.\ \ref{r_x_gamma}). This is usually interpreted as a partially synchrotron self-absorbed jet emission (BK79; \citealt{fb95}). The emission at a given energy is self-absorbed from the base of the jet up to the height of $z\propto \nu^{-1}$, and it is optically thin at higher $z$. The jet becomes optically thin at all $z$ above $E_{\rm t}$. For a magnetic field of $B\propto z^{-1}$, conserving the magnetic energy flux (or for steeper dependencies), the optically-thin emission at $E> E_{\rm t}$ is dominated by the emission from the base of the jet. This follows, e.g., from the synchrotron flux, $F(E) \propto B^{(p+1)/2}$, where $p$ is the index of the electron power law. For $B\propto z^{-1}$, $F(E)\propto z^{-(p+1)/2}$. The number of emitting electrons for a constant-velocity jet per $\ln z$ is $\propto z$. Thus, ${\rm d}F(E)/{\rm d}\ln z\propto z^{-\alpha}$, where $\alpha=(p-1)/2$ is now the index of the optically thin synchrotron radiation. Integrating over $z$, we see that the first logarithmic interval of the jet length measured from its base provides about a half of the total emission for $\alpha=0.5$. Thus, in this Section we adopt a one-zone approximation, with the emission region at $z\pm z/2$ (in the observer's frame) along the jet. The validity of this approximation for optically-thin emission is confirmed by our calculations of the synchrotron spectrum from the entire jet in Section \ref{self-abs}. 

We consider a broken power-law steady-state electron distribution per unit volume,
\begin{equation}
N(\gamma) \simeq K\exp\left(-2\gamma\over \gamma_{\rm m}\right)\times
\cases{\gamma^{-p}, &$1\sim \gamma_0<\gamma\leq \gamma_{\rm b}$;\cr
 \gamma_{\rm b}\gamma^{-p-1}, &$\gamma\geq \gamma_{\rm b}$,\cr}
\label{ngamma2}
\end{equation}
where $K$ is the normalization, $\gamma$ is the Lorentz factor, and the cooling break, $\gamma_{\rm b}$, and the cutoff, $\gamma_{\rm m}$ (assumed $\gg \gamma_{\rm b}$) are discussed and determined in Section \ref{power} below. In Sections \ref{base}--\ref{self-abs}, we assume that the distribution of electrons responsible for the emission in a region around the turnover energy corresponds to $\gamma<\gamma_{\rm b}$, where we can also neglect the high-energy cutoff, i.e., $N(\gamma)=K\gamma^{-p}$.

The synchrotron emission coefficient (i.e., energy production rate per unit volume and per unit solid angle, \citealt{rl79}) from isotropic relativistic electrons in a unit volume at the jet frame for a general form of $N(\gamma)$ can be approximately described as (see, e.g, \citealt{z12}, hereafter Z12),
\begin{equation}
j_{\rm S}(\epsilon)\equiv m_{\rm e} c^2 {\epsilon{\rm d} \dot n_{\rm S}\over {\rm d}\epsilon{\rm d}\Omega}\simeq
{\sigma_{\rm T} c B_{\rm cr}^2 (B/B_{\rm cr})^{1/2} \epsilon^{1/2} \over 48\upi^2 } N\left(\sqrt{\epsilon B_{\rm cr}\over B}\right),
\label{synspecdelta}
\end{equation}
where $B_{\rm cr}={2\upi m_{\rm e}^2 c^3/e h}$ is the critical magnetic field, $h$ is the Planck constant, $e$ is the electron charge, and $\dot n$ denotes a photon production rate per unit volume. The formulation of the synchrotron coefficients in terms of $\sigma_{\rm T}$ and $B_{\rm cr}$ expresses the close correspondence between the synchrotron and Compton processes, discussed in \citet{bg70}. The angle-integrated emissivity per unit volume is $4\upi j_\epsilon$, and the power emitted in all directions by a volume $V$ in the jet frame is $4\upi V j_\epsilon$. (We note that the similar formalism used in Z12 employed rates integrated over the source volume.) For power-law electrons, we have approximately
\begin{equation}
j_{\rm S}(\epsilon)\simeq
{C_1 \sigma_{\rm T}c K B_{\rm cr}^2  \over 48\upi^2}\left(B\over B_{\rm cr}\right)^{{p+1}\over 2} 
\epsilon^{{1-p}\over 2},
\label{synspecpl}
\end{equation}
where $C_1=1$ in the delta-function approximation of equation (\ref{synspecdelta}), whereas averaging over the pitch angle gives (cf.\ \citealt*{jos74})
\begin{equation}
C_1 = {3^{p+4\over 2} \Gamma\left(3p-1\over 12\right) \Gamma\left(3p+19\over 12\right) \Gamma\left(p+1\over 4\right) \over 2^5\upi^{1\over 2}\Gamma\left(p+7\over 4\right)},
\label{c1}
\end{equation}
where $\Gamma$ is the gamma function, and $C_1=1$ for $p=3$. 

The synchrotron self-absorption coefficient averaged over the pitch angle for a power-law electron distribution can be expressed as (cf.\ \citealt{jos74,z12}),
\begin{eqnarray}
\lefteqn{
\alpha_{\rm S}(\epsilon)=  {C_2\upi \sigma_{\rm T} K\over  2\alpha_{\rm f}}  \left(B\over B_{\rm cr}\right)^{p+2\over 2}\!\! \epsilon^{-{p+4\over 2}}
\simeq {C_2 \upi \sigma_{\rm T} \over  2\alpha_{\rm f}} {B\over B_{\rm cr}}\epsilon^{-2} N\left(\sqrt{\epsilon B_{\rm cr}\over B}\right)
,\label{alphas}}\\
\lefteqn{
C_2={3^{p+3\over 2} \Gamma\left(3p+2\over 12\right) \Gamma\left(3p+22\over 12\right) \Gamma\left(p+6\over 4\right)\over 2^4\upi^{1/2} \Gamma\left(p+8\over 4\right)},}
\end{eqnarray}
where $\alpha_{\rm f}$ is the fine-structure constant and $C_2\simeq 0.9996$ for $p=3$. The second equality in equation (\ref{alphas}) gives a monochromatic approximation for $\alpha_{\rm S}$, which is almost accurate for $p=3$. 

The self-Compton emission is approximately given by (e.g., Z12),
\begin{equation}
j_{\rm SC}(\epsilon)\simeq
{\sigma_{\rm T}m_{\rm e}c^3 \epsilon^{1/2}\over 8\upi}\int^{\min(1/\epsilon,\epsilon/\gamma_0^2)}_0 {(n_{\rm S}+n_{\rm SC})(\epsilon_0)\over \epsilon_0^{1/2}} N\left(\sqrt{\epsilon \over \epsilon_0}\right){\rm d}\epsilon_0,
\label{synsc}
\end{equation}
where $n_{\rm S}+n_{\rm SC}$ is the density of the synchrotron and self-Compton photons,
\begin{equation}
(n_{\rm S}+n_{\rm SC})(\epsilon_0)\simeq {4\upi \Theta_{\rm j} \zeta R_{\rm g}\over \epsilon_0 m_{\rm e} c^3} \left[j_{\rm S}(\epsilon_0)+j_{\rm SC}(\epsilon_0)\right].
\label{sdensity}
\end{equation}
Here, we have approximated the synchrotron and self-Compton photons as isotropic in the jet frame, the average photon time in the source as $\Theta_{\rm j} z/c$, and the integration upper limit in equation (\ref{synsc}) accounts for the Thomson limit (assuring that $\epsilon<\gamma$) and $\gamma>1$. Equations (\ref{synsc}--\ref{sdensity}) can be solved iteratively, which yields all orders of Compton scattering. In the cases considered below, the first order dominates. For a synchrotron spectrum above the turnover energy, $\epsilon_{\rm t}$, the integral can be performed analytically for the first-order Compton scattering,
\begin{equation}
j_{\rm SC}(\epsilon)\simeq
{C_1 (\sigma_{\rm T} K B_{\rm cr})^2 c\Theta_{\rm j} \zeta R_{\rm g} \over 96\upi^2 }\left(B\over B_{\rm cr}\right)^{{p+1}\over 2} \!\!
\epsilon^{{1-p}\over 2}
\ln{\min(\epsilon/\gamma_0^2,1/\epsilon)\over \epsilon_{\rm t}}.
\label{ssc_an}
\end{equation}

Then, we calculate the photon energy and the flux in the observer's frame. For the emission of the jet, we have
\begin{equation}
\epsilon={E\over {\cal D}_{\rm j} m_{\rm e} c^2},\quad
{{\rm d}F\over {\rm d}E }={{\cal D}_{\rm j}^2 j(\epsilon) V\over D^2 \Gamma_{\rm j} m_{\rm e}c^2},\quad {\cal D_{\rm j}}={1\over \Gamma_{\rm j}(1-\beta_{\rm j}\cos i)},
\label{ef}
\end{equation}
where $E$ is the observed dimensional photon energy, ${\cal D}_{\rm j}$ is the jet Doppler factor, and the flux transformation to the observed frame is for a steady-state source \citep{s97}. The flux component from the counterjet should be then added, for which the Doppler factor, ${\cal D_{\rm cj}}$, corresponds to $-\beta_{\rm j}$. Hereafter, we take into account the counterjet in numerical calculations, but since the jet emission for our adopted Cyg X-1 parameters ($i=27\degr$, $\beta_{\rm j}=0.6$) is higher than that of the counterjet by $\sim ({\cal D}_{\rm j}/{\cal D}_{\rm cj})^2\simeq 11$, we neglect the counterjet spectral contribution in some flux estimates. Here, the form of ${\rm d}F/ {\rm d}E$ above assumes the energy unit in $F$ and $E$ is the same, whereas they are often assumed to be different (e.g, as erg and eV, respectively), which, however, can be easily accounted for. 

In equation (\ref{ef}), $V$ is in the jet frame. A characteristic volume between $z-z/2$ and $z+z/2$ equals approximately
\begin{equation}
V\simeq \upi \Theta_{\rm j}^2 (\zeta R_{\rm g})^3\Gamma_{\rm j}.
\label{volume}
\end{equation}
This also equals the volume of a cylindrical region with the radius $\Theta_{\rm j}\zeta R_{\rm g}$ and height $z=\zeta R_{\rm g}$. 

The optically-thin synchrotron emission from the considered jet and counterjet regions in the observer's frame is then given by
\begin{equation}
{{\rm d} F_{\rm S}\over {\rm d}E}\simeq
{C_1 \sigma_{\rm T} K (\zeta R_{\rm g})^3 (\Theta_{\rm j}B_{\rm cr})^2 \over 48\upi m_{\rm e} c D^2}\!\left(B\over B_{\rm cr}\right)^{p+1\over 2}\!\! \left(E\over m_{\rm e} c^2\right)^{{1-p\over 2}}\!\!\!
\left({\cal D}_{\rm j}^{p+3\over 2}  \!\!+{\cal D}_{\rm cj}^{p+3\over 2}  \right).
\label{syn_obs}
\end{equation}
Around $E_{\rm t}$, this spectrum joins the optically-thick emission. The optically-thick spectrum of Cyg X-1 shown in Fig.\ \ref{r_x_gamma} with $\alpha\simeq 0$ has the flux of $F_{\rm t}\simeq 20$ cm$^{-2}$ s$^{-1}$ ($\simeq 13$ mJy), which approximately agrees with the fits of R11. (We note that since the $\sim 0.1$ eV region is strongly dominated by the emission of the star and its wind, see Fig.\ \ref{r_x_gamma}, the actual turnover energy may be different than $\sim$0.1 eV.) We normalize then the optically thin synchrotron spectrum to that from the jet base, at $z_0$, as ${\rm d} F_{\rm S}/ {\rm d}E(E_{\rm t})=F_{\rm t}$. Neglecting the counterjet, this yields the normalization of the electron distribution at the base, $K_0$, vs.\ $B_0$, where $B_0$ is the value of $B$ at $z_0$,
\begin{equation}
K_0= {48\upi F_{\rm t} m_{\rm e} c D^2\over C_1\sigma_{\rm T} B_{\rm cr}^2 \Theta_{\rm j}^2 (\zeta_0 R_{\rm g})^3} {\cal D}_{\rm j}^{-{p+3\over 2}} \left(B_0\over B_{\rm cr}\right)^{-{p+1\over 2}}\left(E_{\rm t}\over m_{\rm e} c^2\right)^{p-1\over 2}.
\label{kel}
\end{equation}

Then, a relationship between $B_0$ and $\zeta_0$ at the jet base is provided by the requirement that the synchrotron spectrum is self-absorbed below $E_{\rm t}=\epsilon_{\rm t} {\cal D}_{\rm j} m_{\rm e}c^2$. The absorption optical depth in the observer's direction can be approximated as $\tau_{\rm S} \simeq 2\alpha_{\rm S}(E) \Theta_{\rm j} z_0/{\cal D}_{\rm j}\sin i$, where $2 \Theta_{\rm j} z_0/\sin i$ is the path length going through the jet spine in the observer's frame, and ${\cal D}_{\rm j}\sin i$ is the sine in the jet frame. Taking into account equation (\ref{kel}), we solve $\tau_{\rm S}=1$ at $E_{\rm t}$ for,
\begin{equation}
\zeta_0= \left(48  F_{\rm t} C_2 m_{\rm e}c\over \alpha_{\rm f}C_1 \Theta_{\rm j} \sin i\right)^{1\over 2} {\upi D \over R_{\rm g} B_{\rm cr}} 
\left(B_0\over {\cal D}_{\rm j}B_{\rm cr}\right)^{1\over 4} \left(E_{\rm t}\over m_{\rm e} c^2\right)^{-{5\over 4}}.
\label{z_abs}
\end{equation}
This relationship is only weakly dependent of $p$ through $C_2/C_1$. The inverse relation has $B_0\propto \zeta_0^4$. 

\subsection{The partially self-absorbed jet}
\label{self-abs}

\begin{figure}
\centerline{\includegraphics[width=\columnwidth]{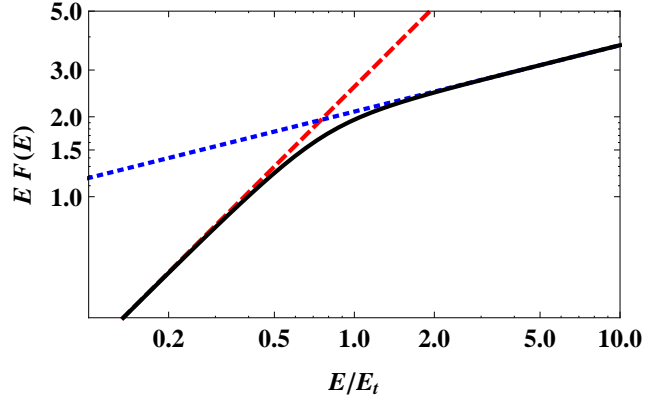}}
\caption{An example of the synchrotron jet spectrum (black solid curve) for $p=2.5$, showing the spectral transition from the optically thick regime to the optically thin one. The normalization corresponds to the integral part of equation (\ref{int2}). The dashed and dotted line shows extrapolation of the optically thick and optically thin regime, respectively.
}
\label{transition}
\end{figure}

We then calculate the synchrotron emission from the entire jet, for a conical jet without energy losses, as assumed by BK79, i.e., with
\begin{equation}
K=K_0 (\zeta/\zeta_0)^{-2},\quad B=B_0(\zeta/\zeta_0)^{-1}.
\label{zeta_dep}
\end{equation}
The source function for synchrotron radiation in the jet frame is,
\begin{equation}
S_{\rm S}(\epsilon)={j_{\rm S}(\epsilon)\over \alpha_{\rm S}(\epsilon)}={C_1 \alpha_{\rm f} c B_{\rm cr}^{5/2} B^{-1/2} \epsilon^{5/2}\over 24\upi^3 C_2}.
\label{source}
\end{equation}
In order to obtain the spectrum observed from the jet, we need to integrate the source function over the projected area of the jet. This implies, cf.\ equation (3) of \citet{heinz06}\footnote{Note that he transformed both the emission and absorption coefficients to the observer's frame by multiplying by ${\cal D}_{\rm j}^2$ ($\delta^2$ in his notation) whereas the latter should be divided by ${\cal D}_{\rm j}$, as well as transformation of the frequency should be taken into account. Consequently, the powers of $\delta$ in subsequent equations need be modified. Also, his emissivity is defined integrated over all directions and the magnetic pressure is assumed to be $B^2/24\upi$.}
\begin{eqnarray}
\lefteqn{
{{\rm d}F_{\rm S}\over {\rm d}E}={\sin i\over m_{\rm e} c^2 D^2} \left[{\cal D}_{\rm j}^3 \int_{z_0}^\infty {\rm d}z\, S_{\rm S} \int_{-\Theta_{\rm j} z}^{\Theta_{\rm j} z}{\rm d}x\left(1-{\rm e}^{-\tau_{\rm S}}\right)+\right.\nonumber}\\
\lefteqn{\quad \left.{\cal D}_{\rm cj}^3 \int_{z_0}^\infty {\rm d}z\, S_{\rm S} \int_{-\Theta_{\rm j} z}^{\Theta_{\rm j} z}{\rm d}x\left(1-{\rm e}^{-\tau_{\rm S}}\right)\right],
\label{int1}}
\end{eqnarray}
where (for the jet)
\begin{equation}
\tau_{\rm S}(\epsilon,z,x)={2\alpha_{\rm S}(\epsilon) \over {\cal D}_{\rm j}\sin i}\left[(\Theta_{\rm j} z)^2-x^2\right]^{1/2}={2\alpha_{\rm S}(\epsilon)\Theta_{\rm j} z \over {\cal D}_{\rm j}\sin i}(1-\psi^2)^{1/2},
\label{tau1}
\end{equation}
$\psi\equiv x/\Theta_{\rm j} z$, and $\epsilon$ needs to be transformed accordingly to the jet and and counterjet formula, respectively. As in Section \ref{base}, we define the turnover energy by $\tau_{\rm S}(\epsilon_{\rm t},z_0,0)=1$, i.e., at the base of the jet for the path going through the spine of the jet. Then, 
\begin{equation}
\tau_{\rm S}=\left({\epsilon\over \epsilon_{\rm t}}{\zeta\over \zeta_0}\right)^{-{p+4\over 2}}(1-\psi^2)^{1\over 2}=\xi^{-{p+4\over 2}}(1-\psi^2)^{1\over 2},
\label{tau2}
\end{equation}
where $\xi\equiv (\zeta/\zeta_0)(\epsilon/\epsilon_{\rm t})$. This allows us to express the flux observed from the jet as
\begin{eqnarray}
\lefteqn{
{{\rm d}F_{\rm S}\over {\rm d}E}={\alpha_{\rm f}C_1 (B_{\rm cr}\zeta_0 R_{\rm g})^2 \Theta_{\rm j} \sin i\over 24\upi^3 C_2 m_{\rm e}c D^2}\left({\cal D}_{\rm j} B_{\rm cr}\over B_0\right)^{1\over 2} \left(E_{\rm t}\over m_{\rm e}c^2\right)^{5\over 2}\nonumber}\\
\lefteqn{\quad\times
\int_{E/E_{\rm t}}^\infty {\rm d}\xi\,\xi^{3\over 2}  \int_{-1}^1 {\rm d}\psi \left\{1-\exp\left[-\xi^{-{p+4\over 2}}\left(1-\psi^2)^{1\over 2}\right)\right]\right\}.}
\label{int2}
\end{eqnarray}
An analogous term for the counterjet (for which $\sin(\upi-i)=\sin i$) needs to be added, including the turnover frequency transformed accordingly. In the optically-thick case, $E\ll E_{\rm t}$, and we can set the lower limit of the outer integration to 0. The resulting dimensionless double integral depends on $p$ only, and we denote it as $C_3$. Its values for $p=2$, 3, 4 are $\simeq 3.61$, 2.10, 1.61, respectively. Then, ${\rm d}F_{\rm S}/ {\rm d}E$ in the optically-thick case has $\alpha=0$,
\begin{equation}
F_{\rm t}={\alpha_{\rm f}C_1 C_3(B_{\rm cr}\zeta_0 R_{\rm g})^2 \Theta_{\rm j} \sin i\over 24\upi^3 C_2 m_{\rm e}c D^2}\left({\cal D}_{\rm j} B_{\rm cr}\over B_0\right)^{1\over 2}\! \left(E_{\rm t}\over m_{\rm e}c^2\right)^{5\over 2}\!\left[1\!+\left({\cal D}_{\rm cj}\over{\cal D}_{\rm j}\right)^{5\over 2}\right],
\label{thick}
\end{equation}
where $E_{\rm t}$ is the jet turnover energy. We can see that this equation times $\upi/(2 C_3)\sim 1$ and neglecting the counterjet contribution is equivalent to equation (\ref{z_abs}). This shows that the one-zone model provides a good approximation for the jet emission at $E_{\rm t}$ (though, of course, the one-zone model does not yield the optically-thick, $\alpha=0$, spectrum). We can also substitute, 
\begin{equation}
\left(E_{\rm t}\over m_{\rm e} c^2\right)^{p+4\over 2} ={\upi C_2 \sigma_{\rm T} K_0\Theta_{\rm j} \zeta_0 R_{\rm g}\over \alpha_{\rm f}\sin i} \left({\cal D}_{\rm j} B_0\over B_{\rm cr}\right)^{p+2\over 2}
\label{et}
\end{equation}
[which follows from $\tau_{\rm S}(\epsilon_{\rm t},z_0,0)=1$] for $E_{\rm t}$ in equation (\ref{int2}) to get the optically-thick flux expressed entirely through the intrinsic jet parameters. We note that, in a dependence opposite to the relativistic beaming along the jet axis of $F_{\rm t}\propto {\cal D}_{\rm j}^{(3p+7)/(p+4)}$ (for the jet), the optically-thick emission is also beamed away from the jet axis in the jet frame, $F_{\rm t}\propto (\sin i)^{(p-1)/(p+4)}$, and $E_{\rm t}\propto {\cal D}_{\rm j}^{(p+2)/(p+4)}/\sin i$. 

Then, for $E\gg E_{\rm t}$, $\tau_{\rm S}(E)\ll 1$, which allows us to perform the double integration in equation (\ref{int2}) analytically, as $(E/E_{\rm t})^{-(p-1)/2}\upi/(p-1)$. We can then substitute $E_{\rm t}$ of equation (\ref{et}) to obtain the flux in the optically-thin case,
\begin{equation}
{{\rm d} F_{\rm S}\over {\rm d}E}=
{C_1 \sigma_{\rm T} K_0 \Theta_{\rm j}^2(\zeta_0 R_{\rm g})^3 B_{\rm cr}^2  \over 24\upi (p-1) m_{\rm e} c D^2}\!\left(B_0\over B_{\rm cr}\right)^{p+1\over 2} \!\!\left(E\over m_{\rm e} c^2\right)^{-{p-1\over 2}}\!\!\!
\left({\cal D}_{\rm j}^{p+3\over 2}  +{\cal D}_{\rm cj}^{p+3\over 2}\right).  \label{syn_thin}
\end{equation}
We see that, for a given $K_0$, this equation times $(p-1)/2=\alpha\sim 1$ equals equation (\ref{syn_obs}), which shows that the one-zone model provides also a good approximation to the optically thin synchrotron emission. Note that the optically-thin emission is (in our approximation to synchrotron) isotropic in the jet frame. 

Only in the transitional region we need to perform the integration numerically. Fig.\ \ref{transition} shows the resulting spectrum (for either jet or counterjet) for $p=2.5$. We see that the transition is rather gradual, not showing a sharp break, in contrast to the broken-power-law fit of R11. We also note that the intersection of the optically thick and thin power laws, ${\rm d} F_{\rm S}/ {\rm d}E({\rm thin})=F_{\rm t}$, occurs not at $E_{\rm t}$ (defined by $\tau_{\rm S}=1$ for the jet) but instead at a $E_{\rm t}^{\rm obs}$, given by
\begin{equation}
{E_{\rm t}^{\rm obs}\over E_{\rm t}}=\left[\upi\over C_3(p-1)\right]^{2\over p-1}.
\label{et2}
\end{equation}
The ratio of equation (\ref{et2}) is $\simeq 0.75$ for $p=2$--5. We take into account this correction to $E_{\rm t}$ in our models in Section \ref{power}. Neglecting the counterjet, the normalization of the electron distribution at the base becomes
\begin{equation}
K_0= {24\upi^2 F_{\rm t} m_{\rm e} c D^2\over C_1 C_3\sigma_{\rm T} B_{\rm cr}^2 \Theta_{\rm j}^2 (\zeta_0 R_{\rm g})^3} \left(B_0\over B_{\rm cr}\right)^{-{p+1\over 2}}\left(E_{\rm t}\over m_{\rm e} c^2\right)^{p-1\over 2}{\cal D}_{\rm j}^{{-p+3\over 2}}.
\label{keljet}
\end{equation}
Note that this equals equation (\ref{kel}) multiplied by $\upi/2 C_3$.

In order to determine the values of $B_0$ and $\zeta_0$ separately, we can assume a degree of equipartition between the pressure or energy density of the magnetic field and the relativistic electrons, as done by \citet*{cdr11}\footnote{Note that our $\beta$ and $\Theta_{\rm j}$ equal their $3\xi$ and $h^{-1}$, respectively. Both the right-hand side of their equation (A2) and the term $m_{\rm e}c/3 e$ should be multiplied by $2\upi$. They neglect the relativistic effects and thus their counterjet flux equals that for the jet.} for the black-hole binary XTE J1550--564. The usual quantity to describe it is the plasma parameter $\beta$, which is the ratio of the particle pressure to that of the field. Since our calculations do not determine the pressure of ions, we define $\beta$ here for the relativistic electrons only,
\begin{equation}
\beta\equiv {u_{\rm e}/3\over B^2/8\upi}={K_0 m_{\rm e} c^2 f/3\over B_0^2/8\upi},
\label{beta}
\end{equation}
where we have assumed the magnetic pressure to be $B^2/8\upi$. A contribution from ions would change the actual value of $\beta$. Here, $u_{\rm e}$ is the energy density of the relativistic electrons, and $f=u_{\rm e}/K m_{\rm e}c^2$, see equation (\ref{ue}) below. Given the dependence of $K$ and $B$ on $z$ of equation (\ref{zeta_dep}), the value of $\beta$ is constant along the jet. We can then substitute $K_0$ of equation (\ref{keljet}) in equation (\ref{et}), and solve it with equation (\ref{thick}) for $B_0$ and $\zeta_0$. Neglecting the counterjet, we find
\begin{eqnarray}
\lefteqn{
B_0=B_{\rm cr} {E_{\rm t}\over m_{\rm e}c^2}{\cal D}_{\rm j}^{-{2p+3\over 2p+13}} \left[(2\alpha_{\rm f}\sin i)^3 C_1 C_3 m_{\rm e} c^3  f^2\over (3 \upi C_2)^3 (\beta B_{\rm cr}\sigma_{\rm T} D)^2 \Theta_{\rm j} F_{\rm t}\right]^{2\over 2p+13},
\label{B0}}\\
\lefteqn{
\zeta_0=\left(E_{\rm t} R_{\rm g}\over m_{\rm e} c^2\right)^{-1}
\left(\upi^3 D^2 F_{\rm t} \over C_1 C_3\right)^{p+6\over 2 p+13} \left(2^3 m_{\rm e} c\over \Theta_{\rm j}B_{\rm cr}^2\right)^{p+7\over 2 p+13}\left(f c\over \beta\sigma_{\rm T}\right)^{1\over 2 p+13}{\cal D}_{\rm j}^{-{p+4\over 2p+13}}\nonumber}\\
\lefteqn{\qquad\times
\left(3 C_2\over \alpha_{\rm f}\sin i\right)^{p+5\over 2 p+13}.
\label{zeta0}}
\end{eqnarray}

Then, we consider the self-Compton component. In the BK79 model, the monochromatic optically-thin synchrotron flux per $\ln \zeta$ is $\propto \zeta^{-(p-1)/2}$. On the other hand, the self-Compton flux in this model decreases with height one power of $\zeta$ faster, $\propto \zeta^{-(p+1)/2}$. This is due to the dependence of the Compton rate on $K^2$, compared to $\propto K$ for synchrotron. Thus, this component is dominated by the base even more than the optically-thin synchrotron. Therefore, instead of solving the self-Compton emission from the entire jet, we consider here only Compton scattering of synchrotron photons produced in the base region. 

The self-Compton component is most important at high energies, where it dominates over the synchrotron one. At those energies, its emission is dominated by the cooled part of the electron distribution (\ref{ngamma2}), with the index $p+1$ at $\gamma>\gamma_{\rm b}$. Neglecting the high-energy cutoff and the counterjet emission (which we take into account in the numerical calculations in Section \ref{power}), the self-Compton flux can be calculated using equations (\ref{ssc_an}--\ref{ef}),
\begin{eqnarray}
\lefteqn{
{{\rm d} F_{\rm SC}\over {\rm d}E}\simeq 
{C_1(\sigma_{\rm T} B_{\rm cr} K_0\gamma_{\rm b})^2 \Theta_{\rm j}^3 (\zeta_0 R_{\rm g})^4 {\cal D}_{\rm j}^{p+4\over 2}\over 
96\upi m_{\rm e}c D^2}
\left(B_0\over B_{\rm cr}\right)^{{p+2\over 2}} \left(E\over m_{\rm e} c^2\right)^{-{p\over 2}} \nonumber}\\
\lefteqn{
\quad\qquad \times \ln\min \left({B_{\rm cr} E/m_{\rm e}c^2\over {\cal D}_{\rm j} B_0\gamma_{\rm b}^2},{{\cal D}_{\rm j} B_{\rm cr} m_{\rm e}c^2 \over E B_0\gamma_{\rm b}^2}\right).}
\label{ssc_obs}
\end{eqnarray}
In the present formalism, we assume the turnover energy corresponds to the un-cooled electrons with the index $p$; then $K_0$ is given by equation (\ref{et}). However, it may also correspond to the cooled electrons, in which case equation (\ref{et}) would need to be modified (by substituting $p+1$ for $p$ and $K_0\gamma_{\rm b}$ for $K_0$). 

The main observational constraint on the amplitude of the self-Compton component comes from observations at $\ga 0.1$ GeV. We use here the \agile\/ upper limit, $F_\gamma(E_\gamma)$, as given in Section \ref{pars}. Since the self-Compton flux decreases with both the increasing magnetic field and the increasing height along the jet, the constraint,
\begin{equation}
{{\rm d} F_{\rm SC}\over {\rm d}E}(E_\gamma)<F_\gamma,
\label{ssc_limit}
\end{equation}
translates into a lower limit on $B_0$ for given $\zeta_0$ and a lower limit on $\zeta_0$ for given $B_0$. The values of $B_0$ and $\zeta_0$ are mutually related by equation (\ref{thick}). Using this, equation (\ref{ssc_limit}) yields the limit for $B_0$,
\begin{equation}
\left(B_0\over B_{\rm cr}\right)^{p+1\over 2} \!\!\!\ga {F_{\rm t}\gamma_{\rm b}^2 \alpha_{\rm f} \sin i \over 4 F_\gamma C_2 C_3{\cal D}_{\rm j}^{{p+1\over 2}}}\! \left(E_{\rm t}\over m_{\rm e} c^2\right)^{2p+3\over 2}\!\!\left(E_\gamma\over m_{\rm e} c^2\right)^{-{p\over 2}}\!\! \ln{{\cal D}_{\rm j} B_{\rm cr} m_{\rm e}c^2 \over E_\gamma B_0\gamma_{\rm b}^2},\label{b_lim}
\end{equation}
independent of $\Theta_{\rm j}$ and $D$. The corresponding limit on $\zeta_0$ can be readily obtained using equations (\ref{ssc_obs}--\ref{thick}). Note that equations (\ref{ssc_obs}) and (\ref{b_lim}) depend on $\gamma_{\rm b}$, which itself depends on $B_0$, and the above limit depends logarithmically on $B_0$. Thus, these equations need to be solved iteratively.

\subsection{The jet parameters}
\label{power}

Given the knowledge of the electron distribution and the magnetic field, we can calculate the components of the jet+counterjet power, 
\begin{eqnarray}
\lefteqn{P_{\rm e}=
{8\upi \over 3} u_{\rm e}\beta_{\rm j} c (\Gamma_{\rm j}\Theta_{\rm j} \zeta R_{\rm g})^2,\label{pe}}\\
\lefteqn{P_{\rm p}=
2\upi \eta_{\rm p} n_{\rm e} m_{\rm p}c^3 \beta_{\rm j}\Gamma_{\rm j} (\Gamma_{\rm j}-1)(\Theta_{\rm j} \zeta R_{\rm g})^2,\label{pp}}\\
\lefteqn{P_B=
{B^2\over 4}\beta_{\rm j} c (\Gamma_{\rm j}\Theta_{\rm j} \zeta R_{\rm g})^2,
\label{pb}}
\end{eqnarray}
where $P_{\rm e}$, $P_{\rm p}$ and $P_B$, is the kinetic power (i.e., the energy flux carried by the jet) in the electrons (including any positrons and taking into account the contribution from pressure), protons (assuming to be cold and not including the rest mass), magnetic field, respectively. Then, $u_{\rm e}$ and $n_{\rm e}$ are the energy density (including the rest mass) and the number density, respectively, of the electrons responsible for the observed emission, and $\eta_{\rm p}$ is the number of protons per emitting electron. Note that $\eta_{\rm p}$ can be $<1$ if there are positrons in the jet, or $>1$ if not all electrons in the jet are accelerated into the distribution (\ref{ngamma2}). The powers of equations (\ref{pe}--\ref{pb}) can be calculated at any point of the jet, and their constancy with $\zeta$ follows from equation (\ref{zeta_dep}). 

In order to calculate $u_{\rm e}$, $n_{\rm e}$, and the synchrotron power, we need to know the full distribution of electrons. The electron adiabatic and synchrotron (neglecting self-absorption) loss rates, and the Lorentz factor at which the two rates are equal are,
\begin{equation}
\dot\gamma_{\rm ad}\simeq {2\beta_{\rm j} \Gamma_{\rm j}c\over 3 z}{\gamma},\quad \dot\gamma_{\rm S}={4\over 3} {\sigma_{\rm T}\over m_{\rm e} c}{B^2\over 8\upi}\gamma^2, \quad \gamma_{\rm S}={4\upi \beta_{\rm j}\Gamma_{\rm j} m_{\rm e}c^2\over B^2 \sigma_{\rm T} \zeta R_{\rm g}},
\label{gad}
\end{equation}
respectively. The factor of $2/3$ in $\dot \gamma_{\rm ad}$ accounts for the expansion being in two dimensions only and we have assumed that the jet is conical. Appendix \ref{losses} considers in more detail the dependence of $\gamma_{\rm S}$ and the corresponding synchrotron photon energy on parameters of accreting sources with jets. We then define the Lorentz factor of electrons emitting at the local turnover energy, $E_{\rm t}(\zeta)$,
\begin{equation}
\gamma_{\rm t}=\left[B_{\rm cr}E_{\rm t}(\zeta) \over B {\cal D}_{\rm j} m_{\rm e}c^2 \right]^{1\over 2}.
\label{gt}
\end{equation}
Note that electrons with $\gamma<\gamma_{\rm t}$, emitting synchrotron emission below $E_{\rm t}$, have a synchrotron loss rate much lower than $\dot\gamma_{\rm S}$, and their adiabatic losses will dominate (neglecting Compton losses). Consequently, the cooling break energy below which adiabatic losses dominate will be approximately at,
\begin{equation}
\gamma_{\rm b}\simeq \max\left(\gamma_{\rm S}, \gamma_{\rm t}\right).
\label{gammab}
\end{equation}
Note that equation (\ref{zeta_dep}) and $E_{\rm t}(\zeta)\propto \zeta^{-1}$ (BK79) imply that $\gamma_{\rm S}\propto\zeta$, $\gamma_{\rm t}\propto\zeta^0$. Here, we neglect this complication and assume a constant value of $\gamma_{\rm b}$ corresponding to the jet base. The steady-state electron distribution above $\gamma_{\rm b}$ has the $p+1$ index.

We assume then a high-energy cutoff is due to radiative loss rate becoming faster than the acceleration rate. Equating $\dot \gamma_{\rm S}$ to the electron acceleration rate occurring on a gyroperiod time scale, $\dot \gamma_{\rm acc}$,
\begin{equation}
\dot \gamma_{\rm acc}={\eta_{\rm acc} e B\over 2\upi m_{\rm e} c},
\label{gacc}
\end{equation}
where $\eta_{\rm acc}\leq 1$ is a scaling factor, the maximum Lorentz factor and the maximum synchrotron energy are obtained as \citep*{gfr83,dj96} 
\begin{equation}
\gamma_{\rm m}^2\simeq {9 B_{\rm cr}\eta_{\rm acc}\over 8\upi\alpha_{\rm f} B},\quad
\epsilon_{\rm m}\simeq {9\eta_{\rm acc} \over 8\upi \alpha_{\rm f}}\simeq 50\eta_{\rm acc}.
\label{emaxs}
\end{equation}
As noted first by \citet{gfr83}, $\epsilon_{\rm m}$ is independent of $B$. In equation (\ref{ngamma2}), we assume e-folding at $\gamma_{\rm m}/2$, which, given our delta function approximation to the synchrotron spectrum, equation (\ref{synspecdelta}), results in an exponential cutoff of the photon distribution at $\epsilon_{\rm m}$. In order to have $\epsilon_{\rm m}\simeq 2$ (neglecting relativistic energy corrections), as fitted in the two-component model to Cyg X-1 (Section \ref{spectral}), $\eta_{\rm acc}\simeq 0.05$ is required, which value we assume. 

Thus, we now adopt the steady-state electron distribution of equation (\ref{ngamma2}), keeping in mind that the actual one may be more complicated. The main complication not taken into account in our presented formalism is the possible presence of a minimum acceleration Lorentz factor $>1$ (see Section \ref{power}). With equation (\ref{ngamma2}), we can calculate $n_{\rm e}$, $u_{\rm e}$ and $P_{\rm S}$, the power of the synchrotron radiation emitted in all directions by the jet and counterjet,
\begin{eqnarray}
\lefteqn{
n_{\rm e}\simeq {K\over p-1}\left(1-{\gamma_{\rm b}^{1-p}\over p}\right),\label{ne}}\\
\lefteqn{
u_{\rm e}=K m_{\rm e} c^2 f,\,\, f=\left[{\gamma_{\rm b}^{2-p}-1\over 2-p}+\gamma_{\rm b}\left(\gamma_{\rm m}\over 2\right)^{1-p}\Gamma\left(1-p,{2\gamma_{\rm b}\over\gamma_{\rm m}}\right) \right], \label{ue}}\\
\lefteqn{
P_{\rm S}\simeq {B_0^2\over 3\upi}\sigma_{\rm T}c K_0 V_0 \left[{\gamma_{\rm b}^{3-p}-\gamma_{\rm t}^{3-p}\over 3-p}+
\left(\gamma_{\rm m}\over 2\right)^{2-p}\Gamma\left(2-p,{2\gamma_{\rm b}\over\gamma_{\rm m}}\right)\right],
\label{psyn}}
\end{eqnarray}
where $V_0$ is the volume, equation (\ref{volume}), at $\zeta_0$, $f$ is the dimensionless energy density, $\Gamma$ is the incomplete gamma function, and we have assumed $\gamma_{\rm m}\gg \gamma_{\rm b}$. In equation (\ref{ne}), we have also neglected the contribution of the high-energy cutoff, and in equation (\ref{psyn}), the emission below $E_{\rm t}$. We note that $P_{\rm S}$ has been integrated over the entire jet length [with the dependencies of equation (\ref{zeta_dep})], but it happens to be equal to the power emitted by the two base volumes in the one-zone approximation. 

Finally, we consider the constraint $P_B=\eta_B P_{\rm j}$, where we use the observational estimate of $P_{\rm j}$, and $\eta_B<1$ from energy conservation. Using equations (\ref{thick}) and (\ref{pb}), we obtain
\begin{equation}\zeta_0<\zeta_{\rm m}
=\left( m_{\rm p} p_{\rm j}\over \beta_{\rm j} \sigma_{\rm T}\right)^{1\over 10} {2^{8\over 5} \upi^{13\over 10} m_{\rm e}^{7\over 5} c^{13\over 5} D^{4\over 5}\over E_{\rm t} B_{\rm cr}\Theta_{\rm j}^{3\over 5}R_{\rm g}^{9\over 10}(\Gamma_{\rm j}{\cal D}_{\rm j})^{{1\over 5}}}
\left(3 F_{\rm t} C_2 \over \alpha_{\rm f}C_1 C_3 \sin i\right)^{2\over 5},
\label{zmax}
\end{equation}
and $\zeta_0=\zeta_{\rm m}\eta_B^{1/10}$. This limit is only weakly depending on $p$ through $C_2/(C_1 C_3)$.

\subsection{Application to Cyg X-1}
\label{cygx1}

\begin{figure*}
\centerline{\includegraphics[width=14cm]{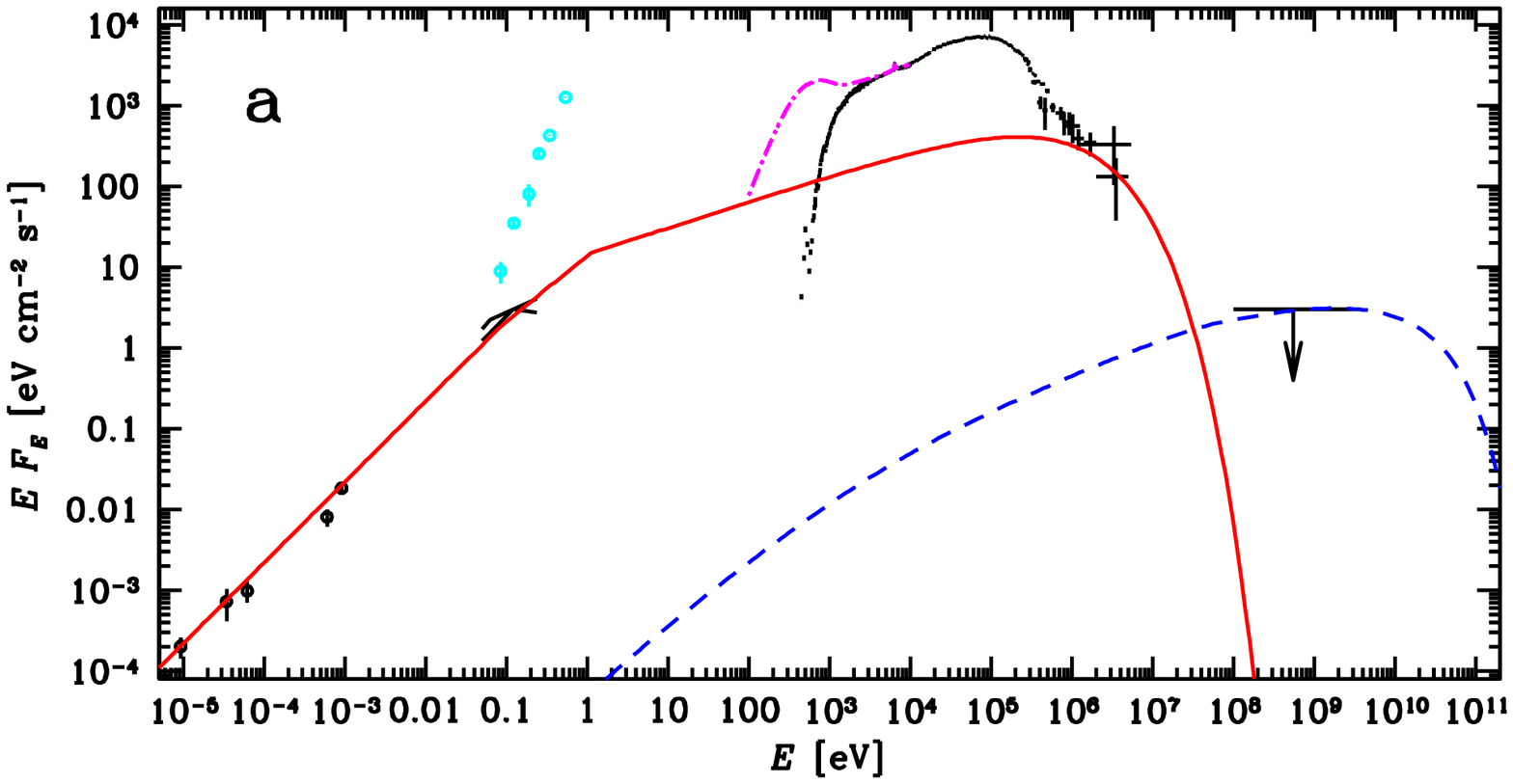}} 
\centerline{\includegraphics[width=14cm]{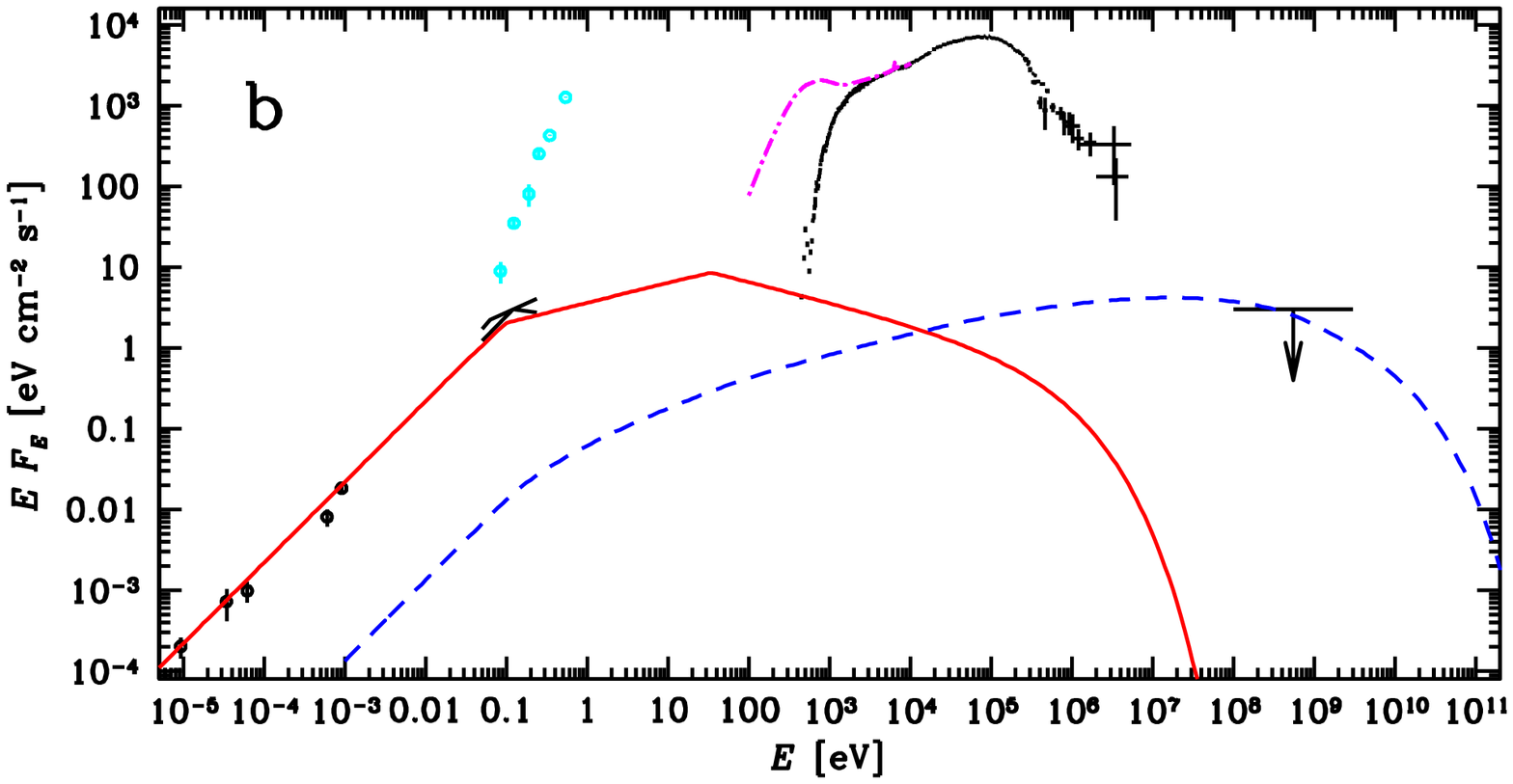}} 
\caption{The data are the same as in Fig.\ \ref{r_x_gamma}. The red solid and blue dashed curves show the model synchrotron and Compton components, respectively. (a) The model 1 for $p=1.35$, accounting for the observed MeV tail, which corresponds to the approximately maximum jet emission allowed by the data. (b) The model 2 for $p=2.5$, in which case the jet emission is well below the MeV tail, with it most likely being emitted by hybrid plasma in the accretion flow. See Section \ref{cygx1} for details.
} \label{model}
\end{figure*}

Here, we consider the full-jet formalism of Sections \ref{self-abs}--\ref{power}. In our model 1, we reproduce the MeV tail by the jet synchrotron emission. Given that the electrons responsible for X/\g-ray emission are cooled, $\gamma>\gamma_{\rm b}$ (see Appendix \ref{losses}), we have first considered a model with $p=1.2$, at which the slope after the cooling brake is 2.2 and the X-ray energy index is $\alpha=0.6$. However, we have found that although the X-ray-emitting electrons do have $\gamma>\gamma_{\rm b}$, there is a hard [$\alpha=(p-1)/2=0.1$] part of the spectrum above $E_{\rm t}$ emitted by $p=1.2$ electrons because $\gamma_{\rm b}\gg \gamma_{\rm t}$. This results in the 1-MeV flux being above that observed. We have thus increased $p$ to a value at which the model would match the 1-MeV flux. Such a model has $p=1.35$, and it is shown in Fig.\ \ref{model}(a). 

We have first imposed equipartition between the electron and magnetic pressure, $\beta=1$, see equations (\ref{B0}--\ref{zeta0}). This model yields the GeV flux somewhat above the observational upper limit, and a slight increase of $B_0$, corresponding to $\beta\simeq 0.8$, gives the GeV flux at that limit, see Fig.\ \ref{model}(a). The model has $\zeta_0\simeq 860$, $B_0\simeq 8.7\times 10^3$ G, $\zeta_{\rm m}\simeq 2100$, $\gamma_{\rm b}=\gamma_{\rm S}\simeq 80>\gamma_{\rm t}\simeq 20$. The low value of $p$ of the model results in a rather slight break at $E_{\rm t}$, from $\alpha=0$ to $\alpha=0.175$, which may be not compatible with the break found by R11. The spectrum steepens to $\alpha=0.675$ only around 1 eV. 

The model has $P_{\rm e}\simeq 8\times 10^{33}$ erg s$^{-1}$, $P_B\simeq 2\times 10^{33}$ erg s$^{-1}$, $P_{\rm p}\simeq 7\eta_{\rm p}\times 10^{34}$ erg s$^{-1}$, $P_{\rm S}\simeq 1\times 10^{35}$ erg s$^{-1}$. We note that $P_{\rm p}/\eta_{\rm p}$ is much less than the estimated jet kinetic power, $P_{\rm j}$, of \citet{russell07}, which requires the presence of $\eta_{\rm p}\ga 10^2$ times more protons than those corresponding to the emitting electrons. Increasing $\zeta_0$ (allowed by the GeV upper limit) decreases $P_{\rm el}$, $P_{\rm p}$ and $\beta$, while it increases $P_B$ and it does not affect $P_{\rm S}$.  

In our model 2, we assume $p=2.5$, motivated by $p\simeq 2.5\pm 0.5$ found above $E_{\rm t}$ in the black-hole binary GX 339--4 \citep{gandhi11}. Imposing equipartition, $\beta=1$, we obtain $\zeta_0\simeq 1.2\times 10^3$, $B_0\simeq 1.1\times 10^4$ G, similar to those found for model 1, and $P_{\rm p}\simeq 3\eta_{\rm p}\times 10^{36}$ erg s$^{-1}$. The resulting GeV flux is $>3$ orders of magnitude below the observational upper limit. The model yielding the GeV flux at the upper limit, shown in Fig.\ \ref{model}(b), has $\zeta_0\simeq 0.8\times 10^3$, $B_0\simeq 3\times 10^3$ G, $\beta\simeq 480$, $\gamma_{\rm S}\simeq 780$, $\gamma_{\rm t}\simeq 48$, and $\zeta_{\rm m}\simeq 2600$. Also, $P_{\rm e}\simeq 3\times 10^{34}$ erg s$^{-1}$, $P_B\simeq 2\times 10^{32}$ erg s$^{-1}$, $P_{\rm p}\simeq 4\eta_{\rm p}\times 10^{37}$ erg s$^{-1}$, $P_{\rm S}\simeq 1\times 10^{34}$ erg s$^{-1}$. The value of $P_{\rm p}$ is only slightly above the range of $P_{\rm j}$ of \citet{russell07}, and it can be within it for a slightly higher value of $B_0$ (yielding a lower GeV flux). We note that the weak magnetic field of this model, with $\beta\gg 1$, is similar to that found in the \g-ray emitting region of the jet in Cyg X-3 (Z12). In both models 1 and 2, the jet in Cyg X-1 appears not dominated by e$^\pm$ pairs. 

We note that the acceleration rate may take place only above certain minimum Lorentz factor, $\gamma_1\gg 1$. In Z12, $\gamma_1\sim 10^3$ was found necessary to explain the \g-ray spectrum of Cyg X-3 from \fermi\/ \citep{fermi}. We have thus also considered a model with $\gamma_1=10^3$. The model 3 accounts for the MeV tail, and it has $p=1.6$. The spectrum has an $\alpha\simeq 0.5$ power law above $E_{\rm t}$ (due to electrons at $<\gamma_1$ which are efficiently synchrotron-cooled), breaking at $\sim 0.5$ keV to $\alpha\simeq 0.8$. Given the relative similarity of the spectrum to that of our model 1, we do not show it. Models with $\gamma_1\gg 1$ and $p>1.6$ are also possible. They do not explain the MeV tail, and are relatively similar to our model 2. 

Our models 1 and 2 have $\gamma_{\rm S}>\gamma_{\rm t}$, which appears compatible with observations of relatively hard optically-thin spectra above $E_{\rm t}$ \citep{cf02,gandhi11,cdr11}. However, if jet magnetic fields are strong enough (which, in case of Cyg X-1, is compatible with its GeV upper limit), $\gamma_{\rm S}<\gamma_{\rm t}$ is possible, which is indeed the case for our model 3. Then, the turnover region will correspond to cooled electrons. In this case, the electron index used in Sections \ref{base}--\ref{self-abs} (which treatment is based on the synchrotron emission and absorption around $E_{\rm t}$) equals $p+1$, see equation (\ref{ngamma2}), which will require certain modifications of our presented formalism. 

\subsection{Irradiation of the jet}
\label{irr}

The jet will be irradiated by the central X-ray source, including an optically-thick accretion disc, and by the star (see Section \ref{pars}). The Doppler factor of the central source with respect to the jet is ${\cal D}_{\rm X}=1/[\Gamma_{\rm j}(1-\beta_{\rm j})]\simeq 2$. The synchrotron losses will dominate over X-ray Compton losses (neglecting Klein-Nishina correction) for,
\begin{equation} 
B>B_{\rm X}\equiv {(2L_{\rm X}/c)^{1/2}\over {\cal D}_{\rm X} \zeta R_{\rm g}}\simeq 6\times 10^3\left(\zeta\over 10^3\right)^{-1}\, {\rm G}.
\label{X}
\end{equation}
Our models have comparable values of $B_0$, and this effect should be taken into account in a more detailed treatment. Note that for the standard dependence of $B\propto z^{-1}$ above the jet base, the ratio between the X-ray Compton and synchrotron losses will be constant along the jet. 

The Compton-scattered emission will be mostly directed back towards the X-ray source due to anisotropy of relativistic Compton scattering, and only a small fraction, a few per cent of the average Compton emission, will be directed towards the observer at $i\simeq 27\degr$ \citep*{ghisellini91,dch10,z12}. The direct emission will contribute to the high-energy soft \g-ray tail observed in Cyg X-1, and the one reflected from the accretion disc will contribute to the observed Compton reflection spectral component. On the other hand, the same kind of emission from the counterjet will be beamed along the jet axis, but it will be blocked by the disc. Its absence may put some constraints on the outer radius of the disc (which may be relatively small in wind-fed accretion, which appears to take place in Cyg X-1).

The magnetic field which energy density equals the stellar photon energy density at the height $z$ along the jet is,
\begin{equation}
B_*={(2L_*/c)^{1/2} \over {\cal D_*} R},\quad {\cal D_*}={1\over \Gamma_{\rm j}(1-\beta_{\rm j}z/R)},\quad R=(z^2+a^2)^{1/2}. 
\label{ustar}
\end{equation}
At  $z\ll a$, $B_*\simeq 110$ G, which implies the stellar irradiation at the jet base is completely negligible. We note, however, that for $B\propto z^{-1}$ above the jet base, the Compton losses on the stellar emission may become dominant at large heights. For our models 1 and 2, the stellar Compton losses overcome the synchrotron losses at $z\simeq 0.05 a$, $0.02 a$, and become 60 and 600, respectively, times stronger than the synchrotron losses at $z\ga a$. Thus, this appears to be a highly important effect, which should be taken into account for an outer jet of Cyg X-1.

\section{Discussion and conclusions}
\label{discussion}

In Section \ref{spectra}, we have calculated the average hard-state spectra from \integral\/ ISGRI and PICsIT detectors over the period of 2003--2010. They are found to agree well with those from \gro\/ (M02). This is consistent with the long-term X-ray and radio behaviour of Cyg X-1 in the hard state being rather constant since at least 1995 \citep{z11b}, and, in hard X-rays at $>20$ keV, since 1991 \citep{z02}. 

We confirm the presence of a high-energy tail in the $\simeq 0.5$--5 MeV range. We find this tail to be very similar to that measured by \gro, but much weaker than that claimed by L11. This has been independently confirmed by \citet{jrm12}, and then by the revised measurement of Ph.\ Laurent (private communication). 

The origin of the tail is not clear. Although we find it is well fitted by hybrid Comptonization (presumably within a hot accretion flow), it can also be due to a separate spectral component, in particular synchrotron jet emission provided it has a high-energy cutoff at $\sim$1 MeV. If the measurement of the strong polarization above 0.4 MeV of L11 is confirmed, the latter interpretation has to hold. We note, however, that the polarization fraction of $0.67\pm 0.30$ (L11) is difficult to explain even by synchrotron models, as it requires extremely well ordered magnetic fields. For example, polarization fraction of synchrotron radiation in blazars never exceeds 50 per cent, it very rarely reaches 40 per cent, and the typical values are of the order of $\sim$10 per cent (see, e.g., \citealt{jorstad07} and references therein). A polarization fraction of only $\simeq$2.4 per cent was observed in the IR emission of the black-hole binary XTE J1550--564 \citep{cdr11}, interpreted as optically-thin synchrotron jet emission. 

We note here that the electric vector position angle (hereafter EVPA) found by L11 is $140\degr\pm 15\degr$. They claim it is at least $100\degr$ away from the position angle of the radio jet in Cyg X-1, implying the polarization is perpendicular to the jet. However, as we noted in Section \ref{pars}, the actual position angle of the jet in Cyg X-1 is $-(17\degr$--$24\degr)$ \citep{stirling01,rushton11}. Then, given that the EVPA is invariant upon adding $\pm 180\degr$ (since the electric field of a photon oscillates, changing its sign), the EVPA of L11 can also be written as $-(25\degr$--$55\degr)$, and thus it is approximately consistent with the observed jet angle. Polarization parallel to the jet can be produced if the magnetic field is dominated by the toroidal component, and/or due to compression of chaotic magnetic field in the internal shocks formed with fronts oriented (quasi-)perpendicularly to the jet axis. 

In order to test whether the MeV tail may be the high-energy end of the optically-thin jet synchrotron emission, we have developed a jet model allowing us to determine the parameters of its source. We noted that the synchrotron power-law spectrum begins at the turnover energy ($E_{\rm t}$), which corresponds to energy at which the base of the jet becomes optically thin. This implies that most of the optically thin emission originates in the base, which we model as a single zone. Combining the expressions for the synchrotron emission and absorption at $E_{\rm t}$, we find that the magnetic field and the height of the jet base satisfy a relation, $B_0\propto z_0^4$. Physically, it follows from the requirement of obtaining a given synchrotron flux, which yields the total number of electrons for a given $B_0$, combined with the requirement of self-absorption at $E_{\rm t}$. The higher up the emission originates, the lower the electron density (for a conical jet) and thus the lower self-absorption optical depth for a given $B_0$, which has to be compensated by an increase of $B_0$. 

We have then developed a formalism for the synchrotron emission of the full jet (reformulating the model of BK79) at any energy. The emission is found by analytical formulae in the optically thick and thin regimes. We have confirmed the validity of the one-zone model for the optically thin emission. Our results give also detailed spectra in the transition region, near $E_{\rm t}$, as an integral. The spectra have a smooth transition between the two regimes, whereas R11 fitted the transition region as a broken power law. This may affect the accuracy of their values of $E_{\rm t}$ and $F(E_{\rm t})$.

We then consider synchrotron self-Compton emission. For a given synchrotron flux at $E_{\rm t}$, a monochromatic flux from this process is $\propto B_0^{-(p+2)/2}$. We use the observational upper flux limit at 0.1--3 GeV from \agile. This gives a lower limit on $B_0$, and using the found relationship between $z_0$ and $B_0$, also on $z_0$. Next, noting that the jet Poynting flux cannot exceed the total jet kinetic power estimated observationally \citep{gallo05,russell07}, we obtain upper limits on $B_0$ and $z_0$. We also calculate $B_0$ and $z_0$ from the assumption of equipartition between the magnetic field and relativistic electrons.

The above results are applied to Cyg X-1, using the values of $E_{\rm t}$ and $F(E_{\rm t})$ of R11. This yields the location of the jet base at $\sim 10^3 R_{\rm g}$, and $B_0\sim 10^4$ G, relatively weakly dependent on the assumed value of $p$. We also find that in order for the observed MeV tail to be due to the jet synchrotron emission, $p\simeq 1.3$--1.6, which are relatively low values for acceleration processes. We note that such acceleration indices are harder than that assumed before in X-ray jet models of black-hole binaries, $p\simeq 2$--2.3 \citep*{markoff03,hs03,mhd03,falcke04,heinz04}. Low values of the acceleration index appear also to be in conflict with observations of optically-thin synchrotron radio spectra, e.g., \citet{mj04}. On the other hand, values of $p>2$ are consistent with standard acceleration models and observations of optically thin synchrotron spectra (e.g., \citealt{bo98,kirk00}). Still, our present understanding of electron acceleration appears insufficient to rule out models with $p<2$. 

We find that models with $p>2$ can have a much higher kinetic jet power, due to a higher number of ions associated with the emitting electrons with a steep spectrum, as well as due to the magnetic field allowed then to be low by the GeV upper limit. Our model 2 has the kinetic power about equal to that determined observationally \citep{russell07}. This class of models, however, have the synchrotron MeV fluxes much below those of the MeV tail, and thus the tail cannot be due to that process. This, and the spectral proximity of the tail to the high energy break of the hot disc emission may favour production of the tail in a hot accretion disc, as predicted by the hybrid Comptonization model (\citealt{pc98}; M02; \citealt{pv09,mb09}). 

These values of $B_0$ and $\zeta_0$ are comparable to those recently obtained for XTE J1550--564 assuming a magnetic-electron equipartition in a one-zone model \citep{cdr11}, and for GX 339--4 \citep{gandhi11}. In these objects, the optically-thin emission above the turnover frequency shows $p\simeq 2$--3 \citep{cf02,gandhi11,cdr11}. Given the necessary cooling break (if those indices correspond to the un-cooled part of the electron distribution) corresponding to an energy between the IR and X-rays, the implied jet contribution to the X-rays is low. Interestingly, GX 339--4 also shows a hard X-ray tail on top of thermal Compton spectrum \citep{wardzinski02,droulans10} similar to that in Cyg X-1, which tail then has to have the origin different than synchrotron emission of the jet.

We note that the location of the jet base in Cyg X-1 at $z_0\sim 10^3 R_{\rm g}\simeq 2\times 10^9$ cm is in agreement with the observed large orbital modulation of the radio emission (which is due to orbital-phase dependent free-free absorption by the stellar wind from the donor), with, e.g., the depth of $\simeq 30$ per cent at 15 GHz, which requires the bulk of the radio flux to be emitted at a distance comparable to the orbital separation, $z\sim a\simeq 3\times 10^{12}$ cm \citep{sz07,zdz12}. If the turnover frequency (emitted by the base) is $\nu_{\rm t}\sim (2$--$3)\times 10^{13}$ Hz (R11), $\nu_{\rm t}/(15\,{\rm GHz})\simeq a/z_0$, in agreement with the prediction of the $z\propto \nu^{-1}$ dependence of BK79. 

On the other hand, both our estimate of the jet-base distance and the strong orbital modulation of the radio emission are not consistent with the interpretation in terms of the one-component BK79 model of $\sim$50 per cent of the 8.4 GHz emission observed to be resolved at the scale of $\sim 10^{14}/\sin i$ cm \citep{stirling01}. This interpretation yields the location of the $\tau=1$ region for 8.4 GHz emission at $z\sim 10^{14}$ cm \citep{heinz06}, which implies the jet base at $z_0\sim 3\times 10^{10}$ cm $\sim 10^4 R_{\rm g}$, an order of magnitude above our estimates and implying virtually no orbital modulation at 8--15 GHz. A likely solution of this problem is an occurrence of a secondary dissipation event at a large radius, see a discussion in \citet{zdz12}. Then the inner jet may still follow the BK79 model while the resolved emission is due to secondary dissipation. This also solves the problem of the Cyg X-1 jet power being much lower than the power inferred from the ring nebula, with the jet power for our model 2 about equal to that of \citet{russell07}, and much higher than the theoretical estimate of \citet{heinz06}. (We note, however, that \citealt{heinz06} did not include in his jet power estimate the power of the ion bulk motion, which we find to be dominant, see Section \ref{power}.) We note that two dissipation regions at very different scales, one emitting IR to \g-rays close to the compact object, and one emitting radio far away, are present in Cyg X-3 (Z12).  

Finally, we point out that electrons in the jet base are subject to substantial irradiation by the central X-ray source. For $z\la 10^3 R_{\rm g}$, its energy losses will be dominated by Compton up-scattering of X-rays. We also point out that although irradiation by the stellar photons is negligible at the jet base, it appears to dominate the synchrotron losses at distances along the jet comparable and larger than the orbital separation.

\section*{ACKNOWLEDGMENTS}

We thank J. Poutanen, F. Yuan and the referee for valuable comments, E. Jourdain for providing us with her published SPI spectrum in numerical form, and, especially, Ph.\ Laurent for his published and updated IBIS spectra and valuable discussion. This research has been supported in part by the Polish NCN grants N N203 581240, N N203 404939, and 362/1/N-INTEGRAL/2008/09/0.

\appendix
\section{Electron energy losses and the jet magnetic field}
\label{losses}

In this work, we have found that the relativistic electrons responsible for optically-thin emission at $E\gg E_{\rm t}$ are efficiently cooled for models satisfying the observational constraints. Here, we consider the effect of synchrotron cooling more generally. The importance of the break due to the radiative losses for the electron and photon distributions in X-ray jet models (e.g., of \citealt{markoff03,falcke04}) was earlier pointed out by \citet{hs03}, and then studied in detail by \citet{heinz04}. However, they have not calculated the possible values of this break as a function of the possible magnetic field in the jet.

The Lorentz factor at which the synchrotron losses equal the adiabatic losses, $\gamma_{\rm S}$, is given by equation (\ref{gad}). Since $B\propto M^{-1/2}$ in most equipartition models \citep{hs03,mhd03}, and $z\propto M$ for a given $\zeta$, $\gamma_{\rm S}$ in those models is independent of $M$. We note that $\gamma/\dot \gamma_{\rm ad}$ is both the time scale for the adiabatic losses and the characteristic time scale a flow element of the jet spends at $\sim z$. Thus, for $\gamma>\gamma_{\rm S}$, the loss process has enough time to significantly steepen the electron distribution. If the electrons are reaccelerated locally, their steady state distribution steepens by unity with respect to the accelerated distribution. If the electrons are only advected from lower heights, a high-energy cutoff develops above $\gamma_{\rm S}$. 

The characteristic observed photon energy corresponding to $\gamma_{\rm S}$, see equation (\ref{gad}) (assuming $\gamma_{\rm S}>\gamma_{\rm t}$), emitted at a height of $\zeta$ is
\begin{equation}
{E_{\rm b}\over m_{\rm e} c^2}=
{16\upi^2 {\cal D_{\rm j}} \beta_{\rm j}^2\Gamma_{\rm j}^2 m_{\rm e}^2 c^4\over B^3 B_{\rm cr} \sigma_{\rm T}^2 (\zeta R_{\rm g})^2},
\label{eb0}
\end{equation}
We scale the magnetic field with respect to the jet power [as in equation (\ref{zmax})],
\begin{equation}
{B^2\over 4}\beta_{\rm j}c(\Gamma_{\rm j}\Theta_{\rm j}\zeta R_{\rm g})^2=\eta_B P_{\rm j}.
\label{bscale}
\end{equation}
This yields the cooling-break energy in the observer's frame of,
\begin{eqnarray}
\lefteqn{{E_{\rm b}\over m_{\rm e} c^2}=
{3^{1/2} {\cal D_{\rm j}}\beta_{\rm j}^{7/2}\Gamma_{\rm j}^5 \Theta_{\rm j}^3 \zeta \over 2^2 \eta_B^{3/2} p_{\rm j}^{3/2} m^{1/2}} \left(m_{\rm e}\over m_{\rm p}\right)^{3/2}\! \left(a_0 c^2\over G\msun\right)^{1/2}\label{eb2}}\\
\lefteqn{\quad
\simeq {1.05\times 10^{-12} {\cal D_{\rm j}} \beta_{\rm j}^{7/2}\Gamma_{\rm j}^5 \Theta_{\rm j}^3\zeta \over \eta_B^{3/2} p_{\rm j}^{3/2} m^{1/2}},
 }
\end{eqnarray}
where $m= M/ \msun$, $a_0=r_{\rm e}/\alpha_{\rm f}^2$ is the Bohr radius, and $r_{\rm e}$ is the classical electron radius. We see that to achieve an un-cooled power law extending to hard X-rays, $E_{\rm b}\sim m_{\rm e} c^2$, in a jet with a moderate or high power requires a very low value of $\eta_B$.

We can also relate the magnetic field in the jet to that in the hot accretion flow, as in the models of \citet{hs03}, \citet{mhd03} and \citet{heinz04}. The hot flow is, most likely, described by some variant of the advection-dominated accretion flow (ADAF, \citealt{ny95}). That model is quite close to that of spherical accretion, and, for the sake of simplicity, we will relate the pressure in the latter model, $P_0=m_{\rm p} n v_{\rm ff}^2$ (where $n$ is the proton density and $v_{\rm ff}$ is the free-fall velocity), to the pressure of the magnetic field, $B/8\upi=b P_0$, where $b$ is a coefficient that includes a hot-flow magnetization, a departure of the pressure in the hot flow from that of spherical accretion, and an overall scaling between the hot flow pressure and the jet pressure. The jet is launched from within a hot-flow radius, $r_{\rm j} R_{\rm g}$. We then assume conservation of the magnetic flux, $B\propto z^{-1}$. This yields,
\begin{equation}
B^2={8 \upi b m_{\rm p} c^4 \dot m\over \sigma_{\rm T} r_{\rm j}^{1/2} G M\zeta^2}.
\label{bd}
\end{equation}
The scale factor, $\eta_B$, implied by equations (\ref{bscale}) and (\ref{bd}) is independent of $\zeta$,
\begin{equation}
\eta_B ={b \beta_{\rm j}\Gamma_{\rm j}^2\Theta_{\rm j}^2\dot m\over 2 r_{\rm j}^{1/2} p_{\rm j}}.
\label{bfactor}
\end{equation}
In order to obtain the break energy for this model, this $\eta_B$ should be inserted into equation (\ref{eb2}). Both cases rule out the presence of an un-cooled jet X-ray spectrum, unless the magnetic flux is extremely small compared to the jet flux. In that case, however, the self-Compton emission would yield very strong \g-ray fluxes, in violation of, e.g., the GeV upper limit for Cyg X-1 (see Section \ref{jet}). 

\label{lastpage}

\end{document}